# Quantum Algorithms of Bio-molecular Solutions for the Clique Problem on a Quantum Computer

Weng-Long Chang[1]

[1]Contact Author: Department of Computer Science and Information Engineering, National Kaohsiung University of Applied Sciences, Kaohsiung City, Taiwan 807-78, Republic of China
E-mail: changwl@cc.kuas.edu.tw

Ting-Ting Ren[2], Mang Feng[3] and Jun Luo[4]

[2, 3, 4]State Key Laboratory of Magnetic Resonance and Atomic and Molecular Physics, Wuhan Institute of Physics and Mathematics, Chinese Academy of Sciences, Wuhan, 430071, People's Republic of China
E-mail: [2]ttren@wipm.ac.cn, [3]mangfeng@wipm.ac.cn, [4]jluo@wipm.ac.cn

Kawuu Weicheng Lin[5]

[5]Department of Computer Science and Information Engineering, National Kaohsiung University of Applied Sciences, Kaohsiung City, Taiwan 807-78, Republic of China
E-mail: linwc@cc.kuas.edu.tw

Minyi Guo[6]

[6]Department of Computer Software, the University of Aizu, Aizu-Wakamatsu City, Fukushima 965-8580, Japan
E-mail: minyi@u-aizu.ac.jp

Lai Chin Lu[7]

[7]Contact Author: Department of Industrial and Management Engineering, National Kaohsiung University of Applied Sciences, Kaohsiung City, Taiwan 807-78, Republic of China
E-mail: rachel@cc.kuas.edu.tw

___

Feynman in 1961 [Feynman 1961] and in 1982 [Feynman 1982] respectively proposed molecular computation and one of the most important problems in computation theory, that is, whether computing devices based on quantum theory are able to finish computations faster than the standard Turing machines [Turing 1936]. In 1994, Adleman [Adleman 1994] succeeded to solve an instance of the Hamiltonian path problem in a test tube, just by handling DNA strands. Deutsch [Deutsch 1985] denoted a general model of quantum computation — the quantum Turing machine. From [Garey and Johnson 1979; Karp 1975; Hopcroft et al. 2001], the *clique* problem is a famous NP-complete problem, and its corresponding DNA-based algorithm has been offered in [Ho et al. 2005]. However, its corresponding quantum algorithm has been very rarely studied before. In this paper, it is demonstrated that the DNA-based algorithm [Ho et al. 2005] for solving an instance of the clique problem to any a graph $G = (V, E)$ with $n$ vertices and $\theta$ edges and its *complementary* graph $\overline{G} = (V, \overline{E})$ with $n$ vertices and $m = (((n * (n - 1)) / 2) - \theta)$ edges can be implemented by Hadamard gates, **NOT** gates, **CNOT** gates, **CCNOT** gates, Grover's operators, and quantum measurements on a quantum computer. It is also demonstrated that if Grover's algorithm is employed to accomplish the readout step in the DNA-based algorithm, the quantum implementation of the DNA-based algorithm is equivalent to the oracle work (in the language of Grover's algorithm), that is, the target state labeling preceding Grover's searching steps. It is shown that one oracle work can be completed with $O((2 * n) * (n + 1) * (n + 2) / 3)$ **NOT** gates, one **CNOT** gate and $O((4 * m) + (((2 * n) * (n + 1) * (n + 14)) / 6))$ **CCNOT** gates. This is to say that for the quantum implementation of the DNA-based algorithm [Ho et al. 2005] a faster labeling of the target state is attained, which also implies a speedy solution to an instance of the clique problem.

## 1. INTRODUCTION

The molecular computation was first proposed in 1961 by Feynman [Feynman 1961], while this idea had not been tested experimentally until 1994 when Adleman successfully solved an instance of the Hamiltonian path problem in a test tube just by handling DNA strands [Adleman 1994]. On the other hand, in 1982 Feynman [Feynman 1982] presented one of the most important problems in computation theory, that is whether quantum mechanically computing devices are able to finish computations faster than the standard Turing machines [Turing 1936]. Following this idea, Benioff [Benioff 1982] wrote something about the possibility of quantum computation, and Deutsch [Deutsch 1985] proposed a general model of quantum computation — the quantum Turing machine.

In this paper, it is first demonstrated that for any NP-complete problem, $2^n$ possible solutions encoded by $2^n$

DNA strands form an orthonormal basis of a Hilbert space (a complex vector space, $C^{2^n}$), where *n* is the number of bits for input size of the problem. Next, it is also demonstrated that on a molecular computer as proposed by Adleman [Adleman 1994], the DNA-based algorithm [Ho et al. 2005] to solve an instance of the clique problem for any a graph *G* = (*V*, *E*) with *n* vertices and $\theta$ edges and its *complementary* graph $\overline{G} = (V, \overline{E})$ with *n* vertices and $m = \frac{n \times (n-1)}{2} - \theta$ edges can be implemented by Hadamard gates, **NOT** gates, **CNOT** gates, **CCNOT** gates, Grover's operators, and quantum measurements on a physical quantum computer as proposed by Deutsch [Deutsch 1985]. It is then also proven that if **Grover's algorithm** is applied to accomplish the readout step in the DNA-based algorithm, the quantum implementation of the DNA-based algorithm is equivalent to the oracle work (in the language of **Grover's algorithm**), that is, the target state labeling preceding Grover's searching steps.

The rest of the paper is organized as follows: in Section 2, the motivation to develop a quantum algorithm for solving an instance of the clique problem to any a graph *G* = (*V*, *E*) with *n* vertices and $\theta$ edges and its *complementary* graph $\overline{G} = (V, \overline{E})$ with *n* vertices and $m = \frac{n \times (n-1)}{2} - \theta$ edges is introduced. In Section 3, the development of quantum computing and molecular computing is illustrated. In Section 4, a high-level description of the proposed quantum algorithm for solving an instance of the clique problem of any a graph *G* = (*V*, *E*) with *n* vertices and $\theta$ edges and its *complementary* graph $\overline{G} = (V, \overline{E})$ with *n* vertices and $m = \frac{n \times (n-1)}{2} - \theta$ edges is given. In Section 5, the time complexity and the space complexity of the proposed quantum algorithm are analyzed for solving an instance of the clique problem to any a graph *G* = (*V*, *E*) with *n* vertices and $\theta$ edges and its *complementary* graph $\overline{G} = (V, \overline{E})$ with *n* vertices and $m = \frac{n \times (n-1)}{2} - \theta$ edges. In Section 6, a brief discussion concludes the paper.

2. MOTIVATION

From [Imre and Balazs 2005], many computing and engineering problems can be traced back to an optimization process which aims to find the extreme value (minimum or maximum point) of an unsorted database or a so-called cost function. Unfortunately, classical solutions suffer from high computational complexity if the database is unsorted or, equivalently, the cost function has many local minimum/maximum points. Proposed quantum computing-based solutions involve the repeated application of Grover's searching algorithm. From [Imre 2007], it was presented how to combine classical binary search with quantum existence testing to design an efficient extreme value searching algorithm for unsorted databases/ cost functions that revealed an alternative approach of currently available best solution.

The clique problem of any a graph *G* = (*V*, *E*) with *n* vertices and $\theta$ edges and its *complementary* graph $\overline{G} = (V, \overline{E})$ with *n* vertices and $m = \frac{n \times (n-1)}{2} - \theta$ edges is a NP-complete problem and is also an optimization problem [Garey and Johnson 1979; Karp 1975; Hopcroft et al. 2001], and its corresponding DNA-based algorithm was offered in [Ho et al. 2005] with $O(n^2 - \theta)$ biological operations, $O(2^n)$ library strands, $O(n)$ tubes and the longest library strand, $O(15*n)$. It will be shown below how to combine the DNA-based algorithm with Grover's searching algorithm to design an extreme value searching algorithm for solving an instance of the clique problem that is an alternative approach of currently available best solution.

3. BRIEF REVIEW OF MOLECULAR COMPUTING AND QUANTUM COMPUTING

The optimal DNA-based algorithm of every NP-complete problem was obtained from its characteristic [Guo et al. 2005]. A potentially significant area of application for DNA algorithms is the breaking of encryption schemes [Chang et al. 2005; Boneh et al. 1996; Adleman et al. 1999; Chang et al. 2004]. Adleman and his co-authors [Braich et al. 2002] had performed experiments to solve a 20-variable 24-clause three-conjunctive normal form (3-CNF) formula. It was recalled from [Kari et al 200 5] for a list of known properties of DNA languages, which are free of certain types of undesirable bonds, and then introduced a general framework in which they can characterize each of these properties by a solution of a uniform formal language inequation. From [Seelig et al. 2006], the analysis of enzyme-free Nucleic Acid to logic circuits was introduced. Moreover, an instance of Knapsack problems has been solved by DNA experiments [Henkel et al. 2007]. It was proposed from [Zhang and Winfree 2008] that an allosteric DNA molecule that, in its active configuration, catalyzes a noncovalent DNA reaction. From [Yin et al. 2008], molecular programs were executed for a variety of dynamic functions: catalytic formation of branched junctions, autocatalytic duplex formation by a cross-catalytic circuit, nucleated dendritic growth of a binary molecular 'tree', and autonomous locomotion of a bipedal walker. It was proposed from [Majumder et al. 2008] for the first known direct derivation of the convergence rates of two and three-dimensional assemblies to equilibrium.

On the other hand, Deutsch [Deutsch 1989] proposed a network-like model, called *quantum computational networks*, and established some of their basic properties. Bernstein and Vazirani [Bernstein and Vazirani 1993] proved that there is an efficient universal quantum Turing machine. Yao [Yao 1993] demonstrated the polynomially equivalence between the quantum circuits by Deutsch [Deutsch 1989] and quantum Turing machines. Afterward, strengths and weaknesses of the quantum Turing machine for solving NP-complete problems were discussed [Bennett et al. 1997]. Then the Deutsch-Jozsa algorithm [Deutsch and Jozsa 1992] and Coppersmith's quantum algorithm for the fast Fourier transform [Coppersmith 1994] were proposed. So far, the most frequently cited quantum algorithms are Shor's algorithms for solving factoring integers and discrete logarithm [Shor 1994] and Grover's search algorithm [Grover 1996] for unsorted databases.

For solving the maximum clique problem a new approach to global optimization and control uses geometric methods and modern quantum mathematics was described from [Yatsenko et al. 2004]. It was demonstrated from [Durr et al. 2004] to the quantum query complexity for minimum spanning tree, graph connectivity, strong graph connectivity and single source shortest paths in the adjacency matrix and in the adjacency list model. It was showed from [Berzina et al. 2004] to quantum query lower bounds for solving the dominating set, the Hamiltonian circuit and the traveling salesman problem. It was proposed from [Hogg 1998] for an efficient algorithm to search in structured database, which were then achieved by NMR experiments [Peng et al. 2002; Zhu et al. 2007]. It was offered from [Magniez et al. 2005] for a quantum query algorithm of finding a triangle. It was presented from [Ambainis and Spalek 2006] for quantum algorithms to determine a maximum matching in a bipartite graph and the network flow problem. Moreover, the lower bounds on quantum query complexity for various problems were studied in [Zhang 2006], showing that efficient quantum algorithms are so hard to design. It was presented from [Doern 2007] to some new quantum query and quantum time algorithms and quantum query complexity bounds for the independent set problem and the graph and subgraph isomorphism problem. From [Ju et al. 2007], it was shown how quantum Boolean circuits can be used to implement the oracle and the inversion-about-average function in Grover's search algorithm. From [Yang et al. 2007], they simulated Grover's searching process under the influence of the cavity decay and showed that their scheme could be achieved efficiently to find a marked state with high reliability. From [Peng et al. 2008], it was proposed for an adiabatic quantum algorithm capable of factorizing numbers, using fewer qubits than Shor's algorithm.

## 4. QUANTUM ALGORITHMS FOR BIO-MOLECULAR SOLUTIONS OF THE CLIQUE PROBLEM

In this section, the definition of the clique problem and the DNA-based algorithm of solving the clique problem [Ho et al. 2005] will be described. Next, from the DNA-based algorithm of solving the clique problem [Ho et al. 2005], its corresponding quantum algorithm is proposed.

### 4.1. DEFINITION OF THE CLIQUE PROBLEM

Assume that $G$ is any a graph and $G = (V, E)$, where $V$ is a set of vertices in $G$ and $E$ is a set of edges in $G$. Also suppose that $V$ is $\{v_1, \ldots, v_n\}$ and $E$ is $\{(v_a, v_b)|\ v_a$ and $v_b$ are, subsequently, elements in $V\}$. Assume that $|V|$ is the number of vertices in $V$ and $|E|$ is the number of edges in $E$. Also assume that $|V|$ is equal to $n$ and $|E|$ is at most $(n \times (n-1))/2$ and is equal to $\theta$ for $1 \leq \theta \leq (n \times (n-1))/2$. Also suppose that for a graph $G$ its *complementary* graph $\overline{G} = (V, \overline{E})$, where $\overline{E}$ is $\{(v_c, v_d)|\ v_c$ and $v_d$ are, respectively, elements in $V$ and $(v_c, v_d)$ is out of $E\}$. Assume that $|\overline{E}|$ is the number of edges in $\overline{G}$ and is equal to $m$ that is equal to $(n * (n - 1)) / 2 - |E|$. Mathematically, a *clique* for a graph $G = (V, E)$ is a *complete* sub-graph to $G$. **Definition 4-1** cited in [Karp 1975; Garey and Johnson 1979] is used to denote the clique problem of graph $G$.

**Definition 4-1**: The clique problem of graph $G$ with $n$ vertices and $\theta$ edges means finding a maximum-sized clique in $G$.

In Figure 4-1, graph $G^1$ consists of two vertices and an edge. This graph denotes such a problem. All of the cliques in $G^1$ are $\{v_2, v_1\}$, $\{v_1\}$, $\{v_2\}$ and $\emptyset$. The maximum-sized clique for $G^1$ is $\{v_2, v_1\}$. Thus, the size of the clique problem in Figure 4-1 is two. It is indicated in [Karp 1975; Garey and Johnson 1979] that finding a maximum-sized clique is a NP-complete problem and is also an optimization problem, so it can be formulated as a "computational search" problem.

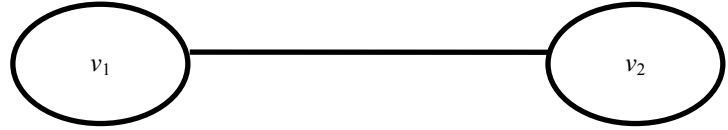

Figure 4-1: the graph $G^1$ of our problem

4.2. ALL OF THE POSSIBLE SOLUTIONS TO THE CLIQUE PROBLEM

It is indicated from **Definition 4-1** that all of the possible solutions to the clique problem of graph $G$ with $n$ vertices and $\theta$ edges consist of $2^n$ possible choices. Every possible choice corresponds to a subset of vertices (a possible clique in $G$). Therefore, assume that $\beta$ is a set of $2^n$ possible choices and $\beta$ is equal to $\{x_n x_{n-1} \ldots x_2 x_1 | \forall x_d \in \{0, 1\}$ for $1 \leq d \leq n\}$. This is to say that the length of each element in $\beta$ is $n$ bits and every element stands for one of $2^n$ possible choices. For the sake of presentation, suppose that $x_d^0$ is used to denote the fact that the value of $x_d$ is zero and $x_d^1$ is applied to denote the fact that the value of $x_d$ is one. If an element $x_n x_{n-1} \ldots x_2 x_1$ in $\beta$ is a legal clique and the value of $x_d$ for $1 \leq d \leq n$ is one, then $x_d^1$ represents that the $d$th vertex is within the legal clique. If an element $x_n x_{n-1} \ldots x_2 x_1$ in $\beta$ is a legal clique and the value of $x_d$ for $1 \leq d \leq n$ is zero, then $x_d^0$ stands for that the $d$th vertex is not within the legal clique. **Definition 4-2** is used to denote how each element in $\beta$ is represented as a unique *computational basis vector* with $2^n$-tuples of binary numbers.

**Definition 4-2**: The $k$th element in $\beta$ can be represented as a unique *computational basis vector* $|u_k\rangle =$

$$\begin{bmatrix} u_{1,1} \\ u_{2,1} \\ \vdots \\ u_{2^n,1} \end{bmatrix}_{2^n \times 1} = \begin{bmatrix} u_{1,1} & u_{1,2} & \cdots & u_{1,2^n} \end{bmatrix}_{1 \times 2^n}^T, \text{ where } u_{k,1} = 1, \text{ and } \forall\ u_{h,1} = 0 \text{ for } 1 \leq h \neq k \leq 2^n.$$

It is pointed out from **Definition 4-2** that the corresponding computational basis vector for the *first* element $x_n^0 x_{n-1}^0 \ldots x_2^0 x_1^0$ in $\beta$ is equal to $\begin{bmatrix} 1 & 0 & \cdots & 0 \end{bmatrix}_{1 \times 2^n}^T$, the corresponding computational basis vector for the *second* element $x_n^0 x_{n-1}^0 \ldots x_2^0 x_1^1$ in $\beta$ is equal to $\begin{bmatrix} 0 & 1 & \cdots & 0 \end{bmatrix}_{1 \times 2^n}^T$, and the corresponding computational basis vector for the *last* element $x_n^1 x_{n-1}^1 \ldots x_2^1 x_1^1$ in $\beta$ is equal to $\begin{bmatrix} 0 & 0 & \cdots & 1 \end{bmatrix}_{1 \times 2^n}^T$. For the sake of presentation, assume that $\alpha$ is the set of corresponding computational basis vectors to each element in $\beta$ and $\alpha =$ $\{[1 \ 0 \ \cdots \ 0]_{1\times 2^n}^T \ [0 \ 1 \ \cdots \ 0]_{1\times 2^n}^T \ \ldots \ [0 \ 0 \ \cdots \ 1]_{1\times 2^n}^T\}$. Since every computational basis vector in $\alpha$ is a coordinated vector from [Debnath and Mikusinski 1990], span $\alpha = C^{2^n}$. It is indicated from [Debnath and Mikusinski 1990], $C^{2^n}$ is a Hilbert space. This implies that set $\alpha$ is an *orthonormal* basis of a Hilbert space.

### 4.3. COMPUTATIONAL STATE SPACE OF MOLECULES FOR THE CLIQUE PROBLEM

The following bio-molecular operations cited from [Adleman 1994] will be used to construct computational state space of molecules for solving the clique problem of graph $G$ with $n$ vertices and $\theta$ edges.

**Definition 4-3**: Given set $\beta = \{x_n x_{n-1} \ldots x_2 x_1 | \forall x_d \in \{0, 1\} \text{ for } 1 \leq d \leq n\}$ and a bit $x_j$, the bio-molecular operation "Append-Head" appends $x_j$ onto the head of every element in set $\beta$. The formal representation is written as **Append-Head**$(\beta, x_j) = \{x_j x_n x_{n-1} \ldots x_2 x_1 | \forall x_d \in \{0, 1\} \text{ for } 1 \leq d \leq n \text{ and } x_j \in \{0, 1\}\}$.

**Definition 4-4**: Given set $\beta = \{x_n x_{n-1} \ldots x_2 x_1 | \forall x_d \in \{0, 1\} \text{ for } 1 \leq d \leq n\}$ and a bit $x_j$, the bio-molecular operation, "Append-Tail", appends $x_j$ onto the end of every element in set $\beta$. The formal representation is written as **Append-Tail**$(\beta, x_j) = \{x_n x_{n-1} \ldots x_2 x_1 x_j | \forall x_d \in \{0, 1\} \text{ for } 1 \leq d \leq n \text{ and } x_j \in \{0, 1\}\}$.

**Definition 4-5**: Given set $\beta = \{x_n x_{n-1} \ldots x_2 x_1 | \forall x_d \in \{0, 1\} \text{ for } 1 \leq d \leq n\}$, the bio-molecular operation "**Discard**$(\beta)$" sets $\beta$ to be an empty set and can be represented as "$\beta = \varnothing$".

**Definition 4-6**: Given set $\beta = \{x_n x_{n-1} \ldots x_2 x_1 | \forall x_d \in \{0, 1\} \text{ for } 1 \leq d \leq n\}$, the bio-molecular operation "**Amplify**$(\beta, \{\beta_i\})$" creates a number of identical copies $\beta_i$ of set $\beta$, and then "Discard$(\beta)$".

**Definition 4-7**: Given set $\beta = \{x_n x_{n-1} \ldots x_2 x_1 | \forall x_d \in \{0, 1\} \text{ for } 1 \leq d \leq n\}$ and a bit, $x_j$, if the value of $x_j$ is equal to one, then the bio-molecular *extract* operation creates two new sets, $+(\beta, x_j^1) = \{x_n x_{n-1} \ldots x_j^1 \ldots x_2 x_1 | \forall x_d \in \{0, 1\} \text{ for } 1 \leq d \neq j \leq n\}$ and $-(\beta, x_j^1) = \{x_n x_{n-1} \ldots x_j^0 \ldots x_2 x_1 | \forall x_d \in \{0, 1\} \text{ for } 1 \leq d \neq j \leq n\}$. Otherwise, it produces another two new sets, $+(\beta, x_j^0) = \{x_n x_{n-1} \ldots x_j^0 \ldots x_2 x_1 | \forall x_d \in \{0, 1\} \text{ for } 1 \leq d \neq j \leq n\}$ and $-(\beta, x_j^0) = \{x_n x_{n-1} \ldots x_j^1 \ldots x_2 x_1 | \forall x_d \in \{0, 1\} \text{ for } 1 \leq d \neq j \leq n\}$.

**Definition 4-8**: Given $m$ sets $\beta_1 \ldots \beta_m$, the bio-molecular *merge* operation, $\cup(\beta_1, \ldots, \beta_m) = \beta_1 \cup \ldots \cup \beta_m$.

**Definition 4-9**: Given set $\beta = \{x_n x_{n-1} \ldots x_2 x_1 | \forall x_d \in \{0, 1\} \text{ for } 1 \leq d \leq n\}$, the bio-molecular operation "**Detect**$(\beta)$" returns *true* if $\beta \neq \varnothing$. Otherwise, it returns *false*.

**Definition 4-10**: Given set $\beta = \{x_n x_{n-1} \ldots x_2 x_1 | \forall x_d \in \{0, 1\} \text{ for } 1 \leq d \leq n\}$, the bio-molecular operation "**Read**$(\beta)$" describes any element in $\beta$. Even if $\beta$ includes many different elements, the bio-molecular operation can give an explicit description of exactly one of them.

For solving the clique problem of a graph with $\theta$ edges and $n$ vertices, the following bio-molecular algorithm can be used to construct all of the $2^n$ possible choices (including legal and illegal cliques). A set $\beta_0$ is an empty set and is regarded as the input set of the following DNA-based algorithm. The second parameter $n$ in **ConstructStateSpace**$(\beta_0, n)$ is applied to stand for the number of vertices.

---

**Procedure ConstructStateSpace**$(\beta_0, n)$
(0a) **Append-Tail**$(\gamma_1, x_n^1)$.
(0b) **Append-Tail**$(\gamma_2, x_n^0)$.
(0c) $\beta_0 = \cup(\gamma_1, \gamma_2)$.
(1) **For** $d = n - 1$ **downto** 1
    (1a) **Amplify**$(\beta_0, \gamma_1, \gamma_2)$.
    (1b) **Append-Tail**$(\gamma_1, x_d^1)$.

---

    (1c) **Append-Tail**($\gamma_2$, $x_d^0$).
    (1d) $\beta_0 = \cup(\gamma_1, \gamma_2)$.
  **End For**
**End Procedure**

**Lemma 4-1**: For solving the clique problem of a graph with $\theta$ edges and $n$ vertices, $2^n$ possible choices constructed from the DNA-based algorithm **ConstructStateSpace**($\beta_0$, $n$) form an orthonormal basis of a Hilbert space (a complex vector space, $C^{2^n}$).

**Proof**:

  The DNA-based algorithm **ConstructStateSpace**($\beta_0$, $n$) is implemented by means of the **amplify**, **append-tail**, and **merge** operations. The first execution of Step (0a) and each execution of Step (0b), respectively, append the value "1" for $x_n$ as the first bit of every element in a set $\gamma_1$ and the value "0" for $x_n$ as the first bit of every element in a set $\gamma_2$. This is to say that $\gamma_1 = \{x_n^1\}$ and $\gamma_2 = \{x_n^0\}$. Next, every execution of Step (0c) generates the set union for the two sets $\gamma_1$ and $\gamma_2$ so that $\beta_0 = \gamma_1 \cup \gamma_2 = \{x_n^1, x_n^0\}$, and $\gamma_1 = \varnothing$ and $\gamma_2 = \varnothing$.

  Each execution of Step (1a) produces two identical copies, $\gamma_1$ and $\gamma_2$, of set $\beta_0$, and $\beta_0 = \varnothing$. Every execution of Step (1b) then appends the value "1" for $x_d$ onto the end of $x_n \ldots x_{d+1}$ for every element in $\gamma_1$. Similarly, each execution of Step (1c) also appends the value "0" for $x_d$ onto the end of $x_n \ldots x_{d+1}$ for every element in $\gamma_2$. Next, each execution of Step (1d) creates the set union for the two sets $\gamma_1$ and $\gamma_2$ so that $\beta_0 = \gamma_1 \cup \gamma_2$, and $\gamma_1 = \varnothing$ and $\gamma_2 = \varnothing$. After repeating to execute Steps (1a) through (1d), $\beta_0 = \{x_n x_{n-1} \ldots x_2 x_1 | \forall x_d \in \{0, 1\} \text{ for } 1 \leq d \leq n\}$. This is to say that $2^n$ possible choices are created and from **Definition 4-2** each possible choice corresponds to a unique *computational basis vector*. Therefore, it is at once inferred that $2^n$ computational state vectors form an orthonormal basis of a Hilbert space (a complex vector space, $C^{2^n}$). ∎

4.4. SPACE OF QUANTUM MECHANICAL SOLUTION FOR THE CLIQUE PROBLEM

  A *qbit* (quantum bit) has two computational basis vectors $|0\rangle$ and $|1\rangle$ of the two-dimensional Hilbert space from [Imre and Balazs 2005; Nielsen and Chuang 2000], and corresponds to the classical bit values 0 and 1. The computational basis vector $|0\rangle$ of a quantum bit is represented as a (2 × 1) column vector $\begin{bmatrix} 1 \\ 0 \end{bmatrix}_{2 \times 1}$ and the computational basis vector $|1\rangle$ of a quantum bit is also represented as a (2 × 1) column vector $\begin{bmatrix} 0 \\ 1 \end{bmatrix}_{2 \times 1}$. It is indicated from [Imre and Balazs 2005; Nielsen and Chuang 2000] that an arbitrary state $|\varphi\rangle$ of a quantum bit is nothing more than a linearly weighted combination of the following computational basis vectors (4.1): $|\varphi\rangle = w_1 |0\rangle + w_2 |1\rangle = w_1 \begin{bmatrix} 1 \\ 0 \end{bmatrix}_{2 \times 1} + w_2 \begin{bmatrix} 0 \\ 1 \end{bmatrix}_{2 \times 1}$, where the weighted factors $w_1$ and $w_2 \in \mathbf{C}^2$ are the so-called probability amplitudes, thus they must satisfy $|w_1|^2 + |w_2|^2 = 1$. A collection of $n$ quantum bits is called a quantum register of size $n$. It may contain any of the $2^n$-dimensional computational basis vectors, $n$ quantum bits of size, or arbitrary superposition of these vectors in [Imre and Balazs 2005; Nielsen and Chuang 2000]. If the content of the quantum bits of a quantum register is known, then the state of the quantum register can be computed by means of a tensor product in the following way: $|\varphi\rangle = \otimes_{d=n}^{1} |\lambda_d\rangle$. It is pointed from [Imre and Balazs 2005; Nielsen and Chuang 2000] that the Hadamard gate $H$ is a quantum gate of one quantum bit (a 2 × 2 matrix), $H_{1,1} = \dfrac{1}{\sqrt{2}}$, $H_{1,2} =$ $\frac{1}{\sqrt{2}}$, $H_{2,1} = \frac{1}{\sqrt{2}}$, and $H_{2,2} = -\frac{1}{\sqrt{2}}$. For a general input (4.1), it produces (4.2): $|\Phi\rangle = H|\varphi\rangle = \frac{l_1+l_2}{\sqrt{2}}|0\rangle + \frac{l_1-l_2}{\sqrt{2}}|1\rangle$. From (4.2), $H|0\rangle = \frac{|0\rangle+|1\rangle}{\sqrt{2}}$, and $H|1\rangle = \frac{|0\rangle-|1\rangle}{\sqrt{2}}$. The generalization of (4.2) for a quantum register with $n$ quantum bits of size in which every individual quantum bit is connected to one Hadamard gate of one quantum bit is the $2^n$ possible choices for the clique problem of any graph with $\theta$ edges and $n$ vertices. For an all-zero input $|\Phi\rangle = \otimes_{d=n}^{1}|0\rangle$, the outcome is (4.3): $|\varphi\rangle = H^{\otimes n}|\Phi\rangle = \frac{1}{\sqrt{2^n}}\sum_{i=0}^{2^n-1}|i\rangle$, where $H^{\otimes n}$ stands for the joined $n$-qbit Hadamard gate. (4.3) is a generalization of (4.2) and is also the computational state space of quantum solutions for solving the clique problem of any a graph with $\theta$ edges and $n$ vertices.

### 4.5. THE DNA-BASED ALGORITHM OF SOLVING THE CLIQUE PROBLEM

The following DNA-based algorithm cited in [Ho et al. 2005] is employed to solve the clique problem of any graph $G$ with $\theta$ edges and $n$ vertices. The notations used in **Algorithm 4-1** are denoted in the previous subsections.

**Algorithm 4-1**: Solving the clique problem of any graph $G = (V, E)$ with $n$ vertices and $\theta$ edges and its *complementary* graph $\overline{G} = (V, \overline{E})$ with $n$ vertices and $m = \frac{n \times (n-1)}{2} - \theta$ edges.

(1) **ConstructStateSpace**($\beta_0$, $n$).
(2) **For** each edge, $e_k = (v_i, v_j)$, in $\overline{G}$ to $1 \le k \le m$ and bits $x_i$ and $x_j$ respectively represent vertices $v_i$ and $v_j$.
    (2a) $\beta_{ON} = +(\beta_0, x_i^1)$ and $\beta_{OFF} = -(\beta_0, x_i^1)$.
    (2b) $\beta_{ON}^1 = +(\beta_{ON}, x_j^1)$ and $\beta_{OFF}^1 = -(\beta_{ON}, x_j^1)$.
    (2c) **Discard**($\beta_{ON}^1$).
    (2d) $\beta_0 = \cup(\beta_{OFF}, \beta_{OFF}^1)$.
  **EndFor**
(3) **For** $i = 0$ **to** $n$-1
    (4) **For** $j = i$ **down to** 0
        (4a) $\beta_{j+1}^{ON} = +(\beta_j, x_{i+1}^1)$ and $\beta_j = -(\beta_j, x_{i+1}^1)$.
        (4b) $\beta_{j+1} = \cup(\beta_{j+1}, \beta_{j+1}^{ON})$.
    **EndFor**
  **EndFor**
(5) **For** $i = n$ **down to** 1
    (5a) **If** (detect($\beta_i$) =="yes") **then**
        (5b) **Read**($\beta_i$) and terminate the algorithm.
    **EndIf**
  **EndFor**
**EndAlgorithm**

The graph in Figure 4-1 includes two vertices and one edge. Hence the values for $n$ and $\theta$ are respectively two and one. For its *complementary* graph, the value of $m$ (the number of its edges) is equal to zero. The execution of Step (1) calls for the algorithm **ConstructStateSpace**($\beta_0$, $n$). The first parameter $\beta_0$ is an empty set and the value of the second parameter is equal to two. The execution of Steps (0a) and (0b) in **ConstructStateSpace**($\beta_0$, $n$) then produces $\gamma_1 = \{x_2^1\}$ and $\gamma_2 = \{x_2^0\}$. The execution of Step (0c) obtains $\beta_0 = \{x_2^1, x_2^0\}$, $\gamma_1 = \varnothing$, and $\gamma_2 = \varnothing$. The first execution of Step (1a) results in $\gamma_1 = \{x_2^1, x_2^0\}$, $\gamma_2 = \{x_2^1, x_2^0\}$ and $\beta_0 = \varnothing$. The first execution of Steps (1b) and (1c) generates $\gamma_1 = \{x_2^1 x_1^1, x_2^0 x_1^1\}$ and $\gamma_2 = \{x_2^1 x_1^0, x_2^0 x_1^0\}$. The first execution of Step (1d) then results in $\beta_0 = \{x_2^1 x_1^1, x_2^0 x_1^1, x_2^1 x_1^0, x_2^0 x_1^0\}$, $\gamma_1 = \varnothing$, and $\gamma_2 = \varnothing$. Therefore, after Step 1 in **Algorithm 4-1** is performed, $\beta_0$ contains $2^2$ possible choices. The first edge in the Figure 4-1 graph is $(v_1, v_2)$. However, there is no edge in its complementary graph. So, Step (2a) through Step (2d) in **Algorithm 4-1** is not executed.

Because the value of *n* is equal to two, Steps (4a) and (4b) in **Algorithm 4-1** will be executed three times. The first execution of Step (4a) in **Algorithm 4-1** results in $\beta_1^{ON} = \{x_2^1 x_1^1, x_2^0 x_1^1\}$ and $\beta_0 = \{x_2^1 x_1^0, x_2^0 x_1^0\}$. The first execution of Step (4b) in **Algorithm 4-1** then results in $\beta_1 = \{x_2^1 x_1^1, x_2^0 x_1^1\}$ and $\beta_1^{ON} = \emptyset$. The second execution of Step (4a) in **Algorithm 4-1** produces $\beta_2^{ON} = \{x_2^1 x_1^1\}$ and $\beta_1 = \{x_2^0 x_1^1\}$. Next, the second execution of Step (4b) in **Algorithm 4-1** generates $\beta_2 = \{x_2^1 x_1^1\}$ and $\beta_2^{ON} = \emptyset$. The third execution of Step (4a) in **Algorithm 4-1** obtains $\beta_1^{ON} = \{x_2^1 x_1^0\}$ and $\beta_0 = \{x_2^0 x_1^0\}$. The third execution of Step (4b) in **Algorithm 4-1** then results in $\beta_1 = \{x_2^1 x_1^0, x_2^0 x_1^1\}$ and $\beta_1^{ON} = \emptyset$. Thus, after each molecular operation in Step (4a) through Step (4b) in **Algorithm 4-1** is completed, it generates $\beta_2 = \{x_2^1 x_1^1\}$ containing one clique, $\{v_2, v_1\}$, $\beta_1 = \{x_2^1 x_1^0, x_2^0 x_1^1\}$ consisting of two cliques, $\{v_2\}$ and $\{v_1\}$, and $\beta_0 = \{x_2^0 x_1^0\}$ including one clique without vertices. Next, in the first execution of Step (5a) in **Algorithm 4-1**, a *true* is returned. Therefore, after the first execution of Step (5b) in **Algorithm 4-1** is performed, the answer is $\{x_2^1 x_1^1\}$ encoding the maximum-sized clique for graph *G* in Figure 4-1 that is $\{v_2, v_1\}$.

**Lemma 4-2**: Using the steps in **Algorithm 4-1**, an instance of the clique problem of any graph *G* with θ edges and *n* vertices can be solved on a molecular computer.

**Proof**: Refer to [Ho et al. 2005].

4.6. INTRODUCTION OF QUANTUM GATES FOR SOLVING THE CLIQUE PROBLEM

The time evolution of the states of quantum registers can be modeled by means of unitary operators, which are often referred to as quantum gates [Imre and Balazs 2005; Nielsen and Chuang 2000]. Hence, a quantum gate can be regarded as an elementary quantum-computing device that completes a fixed unitary operation on selected quantum bits during a fixed period of time. One-qubit and two-qubit quantum gates are *elementary quantum gates*. The **NOT** gate is a one-qubit gate and sets only the (target) bit to its negation. The **CNOT** (*controlled*-**NOT**) gate is a two-qubit gate and flips the second qubit (the target qubit) if and only if the first qubit (the control qubit) is equal to one. The *controlled-controlled*-**NOT** (**CCNOT**) gate is a three-qubit gate and flips the third qubit (the target qubit) if and only if the first and second qubits (the two control qubits) are both one. Graphical representations for **NOT**, **CNOT,** and **CCNOT** are proposed in Figures 4-2, 4-3, and 4-4, respectively. The control quantum bits are graphically represented by a dot, while the target quantum bits are graphically represented by a cross.

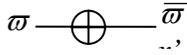

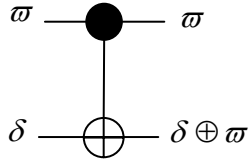

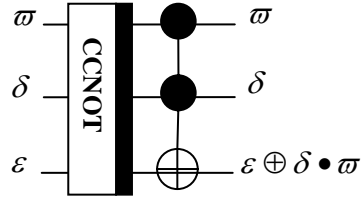

Figure 4-2: **NOT**

Figure 4-3: **CNOT**

Figure 4-4: **CCNOT**

4.7. CONSTRUCTING QUANTUM CIRCUITS FOR FINDING LEGAL CLIQUES

There are $2^n$ possible choices (including legal and illegal cliques) based on the definition of the clique problem for any graph $G = (V, E)$ with *n* vertices and θ edges and its *complementary* graph $\overline{G} = (V, \overline{E})$ with *n* vertices and $m = \dfrac{n \times (n-1)}{2} - \theta$ edges. If any one of $2^n$ possible choices does not include any edge in $\overline{G}$, then it is a legal clique to the question. Otherwise, it is not an answer. Next, among the legal choices, a clique with the maximum size of vertices in *G* needs to be found. Thus, it is supposed that the *k*th edge, $e_k = (v_i, v_j)$, in $\overline{G}$ to $1 \leq k \leq m$ and bits $x_i$ and $x_j$ stand for vertices $v_i$ and $v_j$, subsequently. Since a legal clique does not contain any edge in $\overline{G}$, this is to say that the requested condition can be regarded as finding choices among $2^n$ possible choices that satisfy each formula with the form $\overline{x_i \wedge x_j}$, that is, the *true* value.

Supposed that a quantum register with $n$ quantum bits, $|x_n \cdots x_1\rangle$, is employed to stand for $2^n$ possible choices by means of $n$ **Hadamard** gates operating them. Also assumed that a quantum register of $m$ quantum bit, $|r_m \cdots r_k \cdots r_1\rangle$, for $1 \leq k \leq m$, is used to store the results of evaluating each clause with the form, $\overline{x_i \wedge x_j}$. Suppose that a quantum register of $(m + 1)$ quantum bit, $|c_m c_{m-1} \cdots c_k \cdots c_1 c_0\rangle$, for $0 \leq k \leq m$, is applied to store the results of evaluating the **AND** operation of the current clause (the $k$th clause) and the previous clause (the $(k − 1)$th clause). The operations **NAND** and **AND** are implemented by means of quantum circuits in Figures 4-5 and 4-6, respectively. Therefore, the initial state for each quantum bit in $|r_m \cdots r_k \cdots r_1\rangle$ is prepared in state $|1\rangle$.

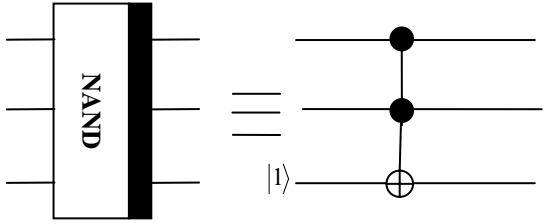 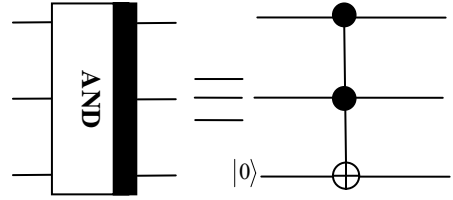

Figure 4-5: **NAND** operation of two Boolean variables.

Figure 4-6: **AND** operation of two Boolean variables.

For evaluating the $k$th clause with the form $\overline{x_i \wedge x_j}$, two quantum registers, $|r_m \cdots r_k \cdots r_1\rangle$ and $|x_n \cdots x_1\rangle$, for $1 \leq k \leq m$ are needed. Hence, its evaluating computation is equal to (4.4):

$$(\otimes_{p=m}^{k+1}|r_p^1\rangle) \otimes (|r_k^1\rangle) \otimes (\otimes_{p=k-1}^{1}|r_p\rangle) \otimes (\otimes_{q=n}^{1}|x_q\rangle) \to (\otimes_{p=m}^{k+1}|r_p^1\rangle) \otimes (|r_k^1 \oplus (x_i \bullet x_j)\rangle) \otimes (\otimes_{p=k-1}^{1}|r_p\rangle) \otimes (\otimes_{q=n}^{1}|x_q\rangle),$$

where $\bullet$ denotes the **AND** operation of two Boolean variables $\{x_i, x_j\}$ for $1 \leq i$ and $j \leq n$, and for $1 \leq p \leq k − 1$, $r_p = r_p^1 \oplus (x_i \bullet x_j)$. The $k$th quantum bit $r_k$ in $|r_m \cdots r_k \cdots r_1\rangle$ is used to store the final result of the evaluating computation for the $k$th clause with the form $\overline{x_i \wedge x_j}$.

The first bit $c_0$ in $|c_m c_{m-1} \cdots c_k \cdots c_1 c_0\rangle$ is initially prepared in state $|1\rangle$, and other $m$ bits in the same quantum register are initially in state $|0\rangle$. Therefore, its evaluating computation which is equivalent to the result of evaluating the **AND** operation of the current clause (the $k$th clause) and the previous clause (the $(k − 1)$th clause) is equal to (4.5):

$$(\otimes_{p=m}^{k+1}|c_p^0\rangle) \otimes (|c_k^0\rangle) \otimes (\otimes_{p=k-1}^{1}|c_p\rangle) \otimes (|c_0^1\rangle) \otimes (\otimes_{p=m}^{k+1}|r_p^1\rangle) \otimes (\otimes_{p=k}^{1}|r_p\rangle) \otimes (\otimes_{q=n}^{1}|x_q\rangle)$$

$$\to (\otimes_{p=m}^{k+1}|c_p^0\rangle) \otimes (|c_k^0 \oplus c_{k-1} \bullet r_k\rangle) \otimes (\otimes_{p=k-1}^{1}|c_p\rangle) \otimes (|c_0^1\rangle) \otimes (\otimes_{p=m}^{k+1}|r_p^1\rangle) \otimes (\otimes_{p=k}^{1}|r_p\rangle) \otimes (\otimes_{q=n}^{1}|x_q\rangle),$$

where $\bullet$ denotes the **AND** operation of two Boolean variables $\{c_{k-1}, r_k\}$ for $1 \leq k \leq m$, and for $1 \leq p \leq k − 1$, $c_p = c_p^0 \oplus c_{p-1} \bullet r_p$. The $k$th quantum bit $c_k$ in $|c_m c_{m-1} \cdots c_k \cdots c_1 c_0\rangle$ is applied to store the result of the evaluating computation for (4.5). This is to say that the $(m + 1)$th quantum bit $c_m$ in the same quantum register is used to store the final result of the evaluating computation for all of the clauses. **Lemma 4-3** is employed to demonstrate how Equations (4.4) and (4.5) perform the parallel logic computation completed by Steps (2a) through (2d) in **Algorithm 4-1** and judge which among the $2^n$ choices are legal cliques and which are not answers.

**Lemma 4-3**: To solve the clique problem of any graph $G = (V, E)$ with $n$ vertices and $\theta$ edges and its *complementary* graph $\overline{G} = (V, \overline{E})$ with $n$ vertices and $m = \dfrac{n \times (n-1)}{2} - \theta$ edges, Equations (4.4) and (4.5) can be employed to complete the parallel logic computation performed by Steps (2a) through (2d) in **Algorithm 4-1** and to judge which among the $2^n$ choices are legal cliques and which are not answers, and quantum evaluating circuits (**QEC**) for implementing the function of Equations (4.4) and (4.5) can be drawn in Figure **4-7**.

**Proof**:

A mathematical induction is applied to carry out the proof. Step (2) in **Algorithm 4-1** is a single loop, and the value of its loop index variable $k$ is applied to represent the $k$th edge in $\overline{G}$ and is from 1 through $m$. When the value of $k$ is equal to one, it is supposed that the first edge $e_1$ in $\overline{G}$ is $(v_i, v_j)$ and bits $x_i$ and $x_j$ stand for vertices $v_i$ and $v_j$, subsequently. The parallel logic computation finished by Steps (2a) through (2b) at the first iteration in the single loop in **Algorithm 4-1** is to select those choices not consisting of the first edge $e_1$. This is to say that the right answers satisfy the formula of the form $\overline{x_i \wedge x_j}$ for $1 \leq i$ and $j \leq n$, that is, the *true* value. Hence, from (4.4), the corresponding evaluating computation is equal to: $(\otimes_{p=m}^{2} |r_p^1\rangle) \otimes (|r_1^1\rangle) \otimes (\otimes_{q=n}^{1} |x_q\rangle) \rightarrow (\otimes_{p=m}^{2} |r_p^1\rangle) \otimes (|r_1^1 \oplus (x_i \bullet x_j)\rangle) \otimes (\otimes_{q=n}^{1} |x_q\rangle)$. The parallel logic computation finished by Steps (2c) through (2d) at the first iteration in the single loop in **Algorithm 4-1** is to reserve legal cliques and discard illegal cliques. This indicates that the reserved choices satisfy the previous clause and the current (first) clause, i.e., the *true* value. Hence, from (4.5), the corresponding evaluating computation is equal to: $(\otimes_{p=m}^{2} |c_p^0\rangle) \otimes (|c_1^0\rangle) \otimes (|c_0^1\rangle) \otimes (\otimes_{p=m}^{2} |r_p^1\rangle) \otimes (|r_1\rangle) \otimes (\otimes_{q=n}^{1} |x_q\rangle) \rightarrow (\otimes_{p=m}^{2} |c_p^0\rangle) \otimes (|c_1^0 \oplus c_0^1 \bullet r_1\rangle) \otimes (|c_0^1\rangle) \otimes (\otimes_{p=m}^{2} |r_p^1\rangle) \otimes (|r_1\rangle) \otimes (\otimes_{q=n}^{1} |x_q\rangle)$.

When the value of $k$ is equal to $a$, it is assumed that the parallel logic computation finished by Steps (2a) through (2b) at the $a$th iteration in the single loop in **Algorithm 4-1** can be completed from (4.4): $(\otimes_{p=m}^{a+1} |r_p^1\rangle) \otimes (|r_a^1\rangle) \otimes (\otimes_{p=a-1}^{1} |r_p\rangle) \otimes (\otimes_{q=n}^{1} |x_q\rangle) \rightarrow (\otimes_{p=m}^{a+1} |r_p^1\rangle) \otimes (|r_a^1 \oplus x_i \bullet x_j\rangle) \otimes (\otimes_{p=a-1}^{1} |r_p\rangle) \otimes (\otimes_{q=n}^{1} |x_q\rangle)$, where for $1 \leq p \leq a - 1$, the value of $r_p$ is equal to the previous evaluating result. It is also supposed that0 the parallel logic computation finished by Steps (2c) through (2d) at the $a$th iteration in the single loop in **Algorithm 4-1** can be completed from (4.5): $(\otimes_{p=m}^{a+1} |c_p^0\rangle) \otimes (|c_a^0\rangle) \otimes (\otimes_{p=a-1}^{1} |c_p\rangle) \otimes (|c_0^1\rangle) \otimes (\otimes_{p=m}^{a+1} |r_p^1\rangle) \otimes (\otimes_{p=a}^{1} |r_p\rangle) \otimes (\otimes_{q=n}^{1} |x_q\rangle) \rightarrow (\otimes_{p=m}^{a+1} |c_p^0\rangle) \otimes (|c_a^0 \oplus c_{a-1} \bullet r_a\rangle) \otimes (\otimes_{p=a-1}^{1} |c_p\rangle) \otimes (|c_0^1\rangle) \otimes (\otimes_{p=m}^{a+1} |r_p^1\rangle) \otimes (\otimes_{p=a}^{1} |r_p\rangle) \otimes (\otimes_{q=n}^{1} |x_q\rangle)$, where for $1 \leq p \leq a - 1$, the value of $c_p$ is equal to the previous evaluating result.

When the value of $k$ is equal to $(a + 1)$, assume that the $(a + 1)$th edge in $\overline{G}$ is $(v_b, v_c)$ and bits $x_b$ and $x_c$ represent vertices $v_b$ and $v_c$, subsequently. The parallel logic computation performed by Steps (2a) through (2b) at the $(a + 1)$th iteration in the single loop in **Algorithm 4-1** is to select those choices not containing the $(a + 1)$th edge. This is to say that the right answers satisfy the formula of the form, $\overline{x_b \wedge x_c}$, that is, the *true* value. Therefore, from (4.4), the corresponding evaluating computation is equal to: $(\otimes_{p=m}^{(a+1)+1} |r_p^1\rangle) \otimes (|r_{a+1}^1\rangle) \otimes (\otimes_{p=(a+1)-1}^{1} |r_p\rangle)$ $\otimes (\otimes^1_{q=n}|x_q\rangle) \rightarrow (\otimes^{(a+1)+1}_{p=m}|r_p^1\rangle) \otimes (|r_{a+1}^1 \oplus x_b \bullet x_c\rangle) \otimes (\otimes^1_{p=(a+1)-1}|r_p\rangle) \otimes (\otimes^1_{q=n}|x_q\rangle)$, where for $1 \leq p \leq (a+1) - 1$, the value of $r_p$ is the previous evaluating result. Next, the parallel logic computation performed by Steps (2c) through (2d) at the $(a+1)$th iteration in the single loop in **Algorithm 4-1** is to reserve legal cliques and to discard illegal cliques. This indicates that the reserved choices satisfy the $a$th clause and the $(a+1)$th clause, i.e., the *true* value. Hence, from (4.5), the corresponding evaluating computation is equal to:

$(\otimes^{(a+1)+1}_{p=m}|c_p^0\rangle) \otimes (|c_{a+1}^0\rangle) \otimes (\otimes^1_{p=a}|c_p\rangle) \otimes (|c_0^1\rangle) \otimes (\otimes^{(a+1)+1}_{p=m}|r_p^1\rangle) \otimes (\otimes^1_{p=a+1}|r_p\rangle) \otimes$
$(\otimes^1_{q=n}|x_q\rangle) \rightarrow (\otimes^{(a+1)+1}_{p=m}|c_p^0\rangle) \otimes (|c_{a+1}^0 \oplus c_a \bullet r_{a+1}\rangle) \otimes (\otimes^1_{p=a}|c_p\rangle) \otimes (|c_0^1\rangle) \otimes (\otimes^{(a+1)+1}_{p=m}|r_p^1\rangle)$
$\otimes (\otimes^1_{p=a+1}|r_p\rangle) \otimes (\otimes^1_{q=n}|x_q\rangle)$, where for $1 \leq p \leq (a+1) - 1$, the value of $c_p$ is equal to the previous evaluating result.

From the statements above, through the relation $(|r_k^1 \oplus x_i \bullet x_j\rangle)$ for $1 \leq i$ and $j \leq n$ and $1 \leq k \leq m$, quantum evaluating circuit (**QEC**) in Figure 4-7 requires figuring out $m$ **NAND** quantum gates for performing the function of Equation (4.4). Next, through the relation $(|c_k^0 \oplus c_{k-1} \bullet r_k\rangle)$ for $1 \leq k \leq m$, quantum evaluating circuit (**QEC**) in Figure 4-7 requires figuring out $m$ **AND** quantum gates for finishing the function of Equation (4.5). Therefore, it is derived that Equations (4.4) and (4.5) can be applied to carry out the parallel logic computation performed by Steps (2a) through (2d) in **Algorithm 4-1** and to judge which among the $2^n$ choices are legal cliques and which are not answers, and quantum evaluating circuits (**QEC**) for implementing the function of Equations (4.4) and (4.5) can be drawn in Figure 4-7. ■

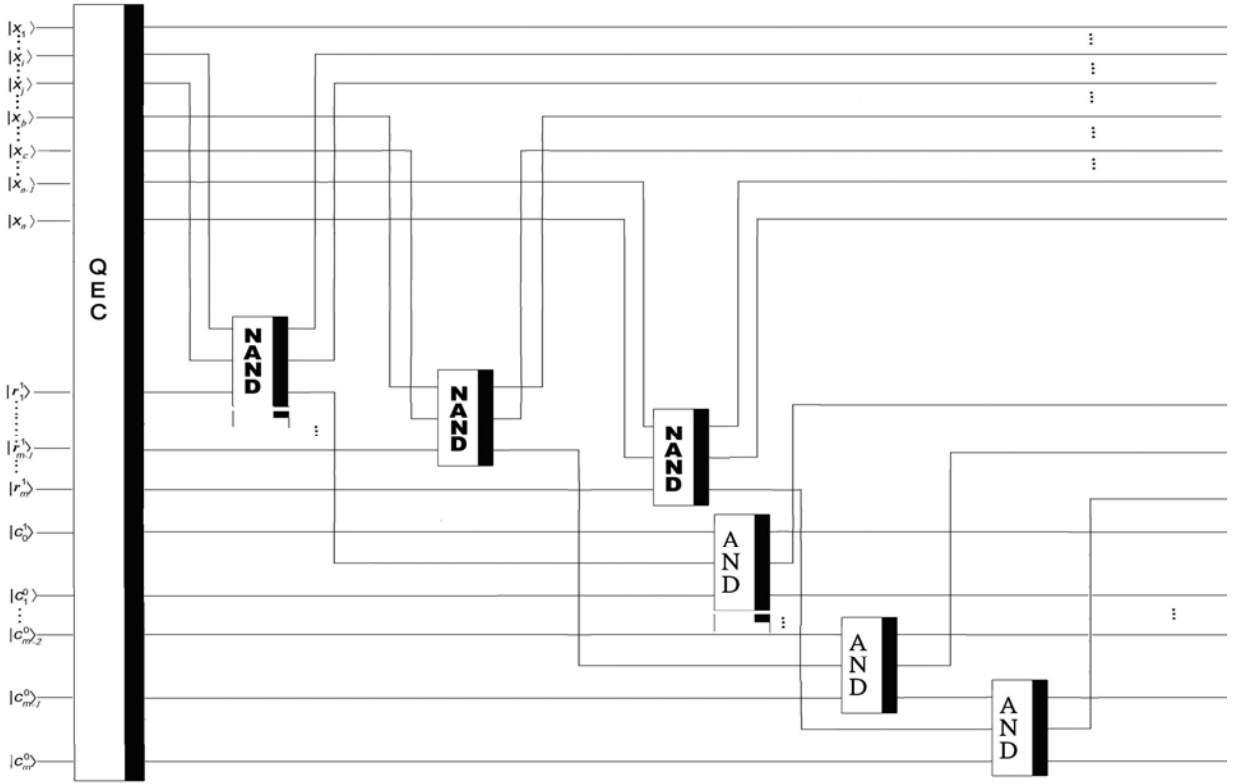

Figure 4-7: Quantum evaluating circuits (**QEC**) for implementing the function of Equations (4.4) and (4.5)

4.8. INTRODUCING QUANTUM NETWORKS FOR FINDING A MAXIMUM-SIZED CLIQUE

The answer of solving the clique problem for any graph $G = (V, E)$ with $n$ vertices and $\theta$ edges and its *complementary* graph $\overline{G} = (V, \overline{E})$ with $n$ vertices and $m = \dfrac{n \times (n-1)}{2} - \theta$ edges is to find a clique with the maximum size of vertices among legal cliques in $G$. Since **Algorithm 4-1** is the DNA-based algorithm for solving the clique problem, Steps (3) and (4) in **Algorithm 4-1** are the outer loop and the inner loop, subsequently, of the nested loop and they are applied to find a maximum-sized clique. Because the value of the outer loop index variable $i$ is from 0 to $n - 1$, Steps (3) and (4) in **Algorithm 4-1** are mainly used to compute the influence of $x_{i+1}$ for the number of ones in tubes (sets) $\beta_0$ through $\beta_{j+1}$ for that the value of $j$ is from $i$ through 0. Since each execution of Step (4a) in **Algorithm 4-1** employs the *extract* operation from tube (set) $\beta_j$ to form two different tubes (sets), $\beta_{j+1}^{ON}$ and $\beta_j$, this indicates that tube (set) $\beta_{j+1}^{ON}$ includes those combinations that have $x_{i+1} = 1$ and tube (set) $\beta_j$ contains those combinations that have $x_{i+1} = 0$. Because those combinations in tube (set) $\beta_j$ have $j$ ones, those combinations in $\beta_{j+1}^{ON}$ have $(j + 1)$ ones. Thus, each execution of Step (4b) in **Algorithm 4-1** employs the *merge* operation to pour tube (set) $\beta_{j+1}^{ON}$ into tube (set) $\beta_{j+1}$. This implies that those combinations in tube (set) $\beta_{j+1}$ have $(j + 1)$ ones. Repeat the execution of Steps (4a) and (4b) in **Algorithm 4-1** until the influence of $x_n$ for the number of ones in tubes (sets) $\beta_0$ through $\beta_n$ is processed. This is to say that those combinations in tube (set) $\beta_i$ for $0 \leq i \leq n$ have $i$ ones.

For carrying out the parallel logic computation generated by Steps (4a) and (4b) at the same iteration in **Algorithm 4-1**, auxiliary quantum bits for $0 \leq i \leq n - 1$ and $0 \leq j \leq i$ $|z_{i+1,j}\rangle$, $|z_{i+1,i+1}\rangle$, $|g_{i,j}\rangle$, $|f_{i,j}\rangle$, $|h_{i,j,i-j+1}\rangle$, $|h_{i,j,0}\rangle$, and $|z_{0,0}\rangle$ are needed. For $0 \leq i \leq n - 1$ and $0 \leq j \leq i$, each quantum bit in $|z_{i+1,j}\rangle$, $|z_{i+1,i+1}\rangle$, $|g_{i,j}\rangle$, $|f_{i,j}\rangle$, and $|h_{i,j,i-j+1}\rangle$ is initially prepared in state $|0\rangle$, and each quantum bit in $|h_{i,j,0}\rangle$ and $|z_{0,0}\rangle$ is initially prepared in state $|1\rangle$. Assume that for $0 \leq i \leq n - 1$ and $0 \leq j \leq i$, $|z_{i+1,j+1}\rangle$ is applied to record the status of tube (set) $\beta_{j+1}$ that has $(j + 1)$ ones, and $|z_{i+1,j}\rangle$ is used to record the status of tube (set) $\beta_j$ that has $j$ ones after the influence of $x_{i+1}$ to the number of ones is figured out from the loop iteration $(i, j)$ in the two-level nested loop in **Algorithm 4-1**. **Lemma 4-4** is used to describe that the parallel logic computation is finished by Steps (4a) and (4b) at the same iteration in **Algorithm 4-1**.

**Lemma 4-4**: The parallel logic computation performed by Steps (4a) and (4b) at the same iteration $(i, j)$ in the two-level nested loop in **Algorithm 4-1** is (4.6):

$|z_{i+1,j+1}\rangle = |z_{i+1,j+1}\rangle \oplus (|c_m\rangle \wedge (|x_{i+1}\rangle \wedge |z_{i,j}\rangle \wedge (\wedge_{k=j+2}^{i+1} |\overline{z}_{i+1,k}\rangle)))$ and $|z_{i+1,j}\rangle = |z_{i+1,j}\rangle \oplus (|c_m\rangle \wedge |\overline{x}_{i+1}\rangle \wedge |z_{i,j}\rangle)$.

**Proof**:

A mathematical induction is used to complete the proof. In the two-level nested loop in **Algorithm 4-1**, the value of the *first* loop index variable $i$ is from 0 through $n - 1$, and the value of the *second* loop index variable $j$ is from $i$ down to 0. When the value of $i$ is equal to zero, the value of $j$ is only equal to zero. Thus, the parallel logic computation performed by Steps (4a) and (4b) at the same iteration (0, 0) in the two-level nested loop in **Algorithm 4-1** consists of solutions (legal cliques) in tube (set) $\beta_1$ including the first vertex and solutions (legal cliques) in tube (set) $\beta_0$ not containing the first vertex. From the first and the second conditions of (4.6) and the iteration (0, 0), $|z_{1,1}\rangle = |z_{1,1}\rangle \oplus (|c_m\rangle \wedge |x_1\rangle \wedge |z_{0,0}\rangle)$ and $|z_{1,0}\rangle = |z_{1,0}\rangle \oplus (|c_m\rangle \wedge |\overline{x}_1\rangle \wedge |z_{0,0}\rangle)$ are obtained. Because the initial state for $|z_{1,1}\rangle$ is zero, the value of $|z_{1,1}\rangle$ is changed to one if and only if the values of three quantum bits ($|c_m\rangle$, $|x_1\rangle$ and $|z_{0,0}\rangle$) are all one. Because the initial state for $|z_{1,0}\rangle$ is also zero, the value of $|z_{1,0}\rangle$ is changed to one if and only if the values of three quantum bits ($|c_m\rangle$, $|\bar{x}_1\rangle$ and $|z_{0,0}\rangle$) are all one. This is to say that if the value for $|z_{1,1}\rangle$ is equal to one, then $|z_{1,1}^1\rangle$ is applied to indicate that solutions (legal cliques) in tube (set) $\beta_1$ consist of the first vertex and have one ones, and if the value for $|z_{1,0}\rangle$ is equal to one, then $|z_{1,0}^1\rangle$ is applied to indicate that solutions (legal vertex covers) in tube (set) $\beta_0$ do not consist of the first vertex and have zero ones. Therefore, it is pointed out that the parallel logic computation completed by Steps (4a) and (4b) at the same iteration (0, 0) in the two-level nested loop in **Algorithm 4-1** can be implemented by the first and the second conditions of (4.6).

Next, when the value of $i$ is equal to $p$ for $0 \leq p \leq n - 1$ and the value of $j$ is equal to $p$, assume that the parallel logic computation performed by Steps (4a) and (4b) at iteration $(p, p)$ in the two-level nested loop in **Algorithm 4-1** can be implemented by the first and second conditions of (4.6). When the value of $i$ is equal to $p$, the value of $j$ is then $p - 1$. Therefore, the parallel logic computation performed by Steps (4a) and (4b) at iteration $(p, p - 1)$ in the two-level nested loop in **Algorithm 4-1** contains solutions (legal cliques) in tube (set) $\beta_{(p-1)+1}$ containing the $(p + 1)$th vertex and have $p$ ones, and solutions (legal cliques) in tube (set) $\beta_{p-1}$ do not contain the $(p + 1)$th vertex and have $(p - 1)$ ones. From the first and second conditions of (4.6) and the iteration $(p, p - 1)$ in the two-level nested loop in **Algorithm 4-1**, $|z_{p+1,(p-1)+1}\rangle = |z_{p+1,(p-1)+1}\rangle \oplus (|c_m\rangle \wedge |x_{p+1}\rangle \wedge |z_{p,p-1}\rangle \wedge |\bar{z}_{p+1,p+1}\rangle)$ and $|z_{p+1,p-1}\rangle = |z_{p+1,p-1}\rangle \oplus (|c_m\rangle \wedge |\bar{x}_{p+1}\rangle \wedge |z_{p,p-1}\rangle)$ are obtained.

The initial state for $|z_{p+1,(p-1)+1}\rangle$ is zero, and if its value is changed to one at the previous iteration $(p, p)$ in the two-level nested loop in **Algorithm 4-1**, then the value for $|\bar{x}_{p+1}\rangle$ is one and the value to $|x_{p+1}\rangle$ is zero. Therefore, under this condition, the value one of $|z_{p+1,(p-1)+1}\rangle$ is preserved because the value of $|x_{p+1}\rangle$ is zero. In other words, when the value for $|z_{p+1,(p-1)+1}\rangle$ is not changed to one at the previous iteration $(p, p)$ in the two-level nested loop in **Algorithm 4-1**, the value of $|z_{p+1,(p-1)+1}\rangle$ is changed to one if and only if the values of quantum bits ($|c_m\rangle$, $|x_{p+1}\rangle$, $|z_{p,p-1}\rangle$ and $|\bar{z}_{p+1,p+1}\rangle$) are all one. The value of $|z_{p+1,p-1}\rangle$ is changed to one if and only if the values of three quantum bits ($|c_m\rangle$, $|\bar{x}_{p+1}\rangle$ and $|z_{p,p-1}\rangle$) are all one. This implies that if the value for $|z_{p+1,(p-1)+1}\rangle$ is equal to one, then $|z_{p+1,(p-1)+1}^1\rangle$ is applied to indicate that solutions (legal cliques) in tube (set) $\beta_{(p-1)+1}$ contain the $(p + 1)$th vertex and have $p$ ones, and if the value for $|z_{p+1,p-1}\rangle$ is equal to one, then $|z_{p+1,p-1}^1\rangle$ is applied to indicate that solutions (legal cliques) in tube (set) $\beta_{p-1}$ do not consist of the $(p + 1)$th vertex and have $(p - 1)$ ones. Therefore, it is indicated that the parallel logic computation performed by Steps (4a) and (4b) at iteration $(p, p - 1)$ in the two-level nested loop in **Algorithm 4-1** can be implemented by the first and second conditions of (4.6). Hence, it can be inferred that the parallel logic computation performed by Steps (4a) and (4b) at iteration $(i, j)$ in the two-level nested loop in **Algorithm 4-1** is (4.6): $|z_{i+1,j+1}\rangle = |z_{i+1,j+1}\rangle \oplus (|c_m\rangle \wedge (|x_{i+1}\rangle \wedge |z_{i,j}\rangle \wedge (\wedge_{k=j+2}^{i+1} |\bar{z}_{i+1,k}\rangle)))$ and $|z_{i+1,j}\rangle = |z_{i+1,j}\rangle \oplus (|c_m\rangle \wedge |\bar{x}_{i+1}\rangle \wedge |z_{i,j}\rangle)$. ■

4.9. CONSTRUCTING QUANTUM NETWORKS FOR FINDING A MAXIMUM-SIZED CLIQUE

From **Lemma 4-4**, Equation (4.6) can be applied to figure out which legal cliques own the maximum size of vertices. Therefore, **Lemma 4-5** is used to show how **FMNO** (finding the maximum number of ones), that is complete quantum networks (**CQN**) of Equation (4.6), computing the influence of $x_{i+1}$ to the number of ones among legal cliques, are implemented by means of **NOT** gates, **AND** gates and **CCNOT** gates.

**Lemma 4-5**: **FMNO** (finding the maximum number of ones) that is complete quantum networks (**CQN**) of implementing the function of Equation (4.6) can be implemented by means of **NOT** gates, **AND** gates and **CCNOT** gates, and is drawn in Figure 4-8.

**Proof**:

The first condition of Equation (4.6) in **Lemma 4-4** is $|z_{i+1,j+1}\rangle = |z_{i+1,j+1}\rangle \oplus (|c_m\rangle \wedge (|x_{i+1}\rangle \wedge |z_{i,j}\rangle \wedge (\wedge_{k=j+2}^{i+1} |\overline{z}_{i+1,k}\rangle)))$. Hence, performing the first condition of Equation (4.6) is to compute $|z_{i+1,j+1}\rangle$ for $0 \le i \le n-1$ and $0 \le j \le i$. The task requires figuring out $(i-j)$ **NOT** gates on $\overline{z}_{i+1,k}$ for $j+2 \le k \le i+1$, and $(i-j+2)$ **AND** gates through (1) the relation $h_{i,j,a} \leftarrow h_{i,j,a} \oplus (h_{i,j,a-1} \bullet \overline{z}_{i+1,k})$, and for $1 \le a \le i-j$ and $j+2 \le k \le i+1$, (2) the relation $h_{i,j,i-j+1} \leftarrow h_{i,j,i-j+1} \oplus (h_{i,j,i-j} \bullet z_{i,j})$ and (3) the relation $f_{i,j} \leftarrow f_{i,j} \oplus (h_{i,j,i-j+1} \bullet x_{i+1})$. Next, the task requires computing one **CCNOT** gates through the relation $z_{i+1,j+1} \leftarrow z_{i+1,j+1} \oplus (c_m \bullet f_{i,j})$. The quantum bit $|z_{i+1,j+1}\rangle$ is used to store the evaluating result of the first condition of Equation (4.6) in **Lemma 4-4**. Subsequently, the **NOT** operations on $\overline{z}_{i+1,k}$ for $j+2 \le k \le i+1$ are reversible to restore each quantum bit $\overline{z}_{i+1,k}$ to its previous state. This enables to reuse $\overline{z}_{i+1,k}$ for $j+2 \le k \le i+1$.

Because the second condition of Equation (4.6) in **Lemma 4-4** is $|z_{i+1,j}\rangle = |z_{i+1,j}\rangle \oplus (|c_m\rangle \wedge |\overline{x}_{i+1}\rangle \wedge |z_{i,j}\rangle)$, completing the second condition of (4.6) in **Lemma 4-4** is to figure out $|z_{i+1,j}\rangle$ for $0 \le i \le n-1$ and $0 \le j \le i$. The task requires computing one **NOT** gate on $|x_{i+1}\rangle$ ($|\overline{x}_{i+1}\rangle$), one **AND** gate, and one **CCNOT** gate through the relation $g_{i,j} \leftarrow g_{i,j} \oplus (z_{i,j} \bullet \overline{x}_{i+1})$ and the relation $z_{i+1,j} \leftarrow z_{i+1,j} \oplus (c_m \bullet g_{i,j})$. Subsequently, all **NOT** gates on $|x_n \cdots x_1\rangle$ are reversible to restore every quantum bit in $|x_n \cdots x_1\rangle$ to its superposition state. This enables to preserve the superposition in $|x_n \cdots x_1\rangle$ and to reuse the superposition in $|x_n \cdots x_1\rangle$.

Based on the rules above, **FMNO** (finding the maximum number of ones) that is complete quantum networks (**CQN**) of implementing the function of Equation (4.6) is drawn in Figure 4-9. Thus, it is at once concluded that **FMNO** (finding the maximum number of ones) that is complete quantum networks (**CQN**) of implementing the function of Equation (4.6) can be implemented by means of **NOT** gates, **AND** gates and **CCNOT** gates, and is drawn in Figure 4-8. ∎

4.10. QUANTUM ALGORITHMS FOR COMPUTING THE NUMBER OF ONES TO LEGAL CLIQUES FOR ANY GRAPH WITH θ EDGES AND *n* VERTICES

The following quantum algorithm is presented to work on the physical quantum computer as proposed by Deutsch [Deutsch 1985] and is employed to compute the number of ones to legal cliques for any graph $G$ with θ edges and *n* vertices. For convenience of presentation, suppose that $|x_b^1\rangle$, $|r_k^1\rangle$, $|c_k^1\rangle$, $|h_{i,j,a}^1\rangle$, $|f_{i,j}^1\rangle$, $|g_{i,j}^1\rangle$, $|z_{0,0}^1\rangle$, $|z_{i+1,j}^1\rangle$, and $|z_{i+1,i+1}^1\rangle$ for $1 \le b \le n$, $0 \le k \le m$, $0 \le i \le n-1$, $0 \le j \le i$, and $0 \le a \le i-j+1$, subsequently, denote the value of their corresponding quantum bits to be 1. Also suppopse that $|x_b^0\rangle$, $|r_k^0\rangle$,

$|c_k^0\rangle$, $|h_{i,j,a}^0\rangle$, $|f_{i,j}^0\rangle$, $|g_{i,j}^0\rangle$, $|z_{0,0}{}^0\rangle$, $|z_{i+1,j}{}^0\rangle$, and $|z_{i+1,i+1}{}^0\rangle$ for $1 \leq b \leq n$, $0 \leq k \leq m$, $0 \leq i \leq n-1$, $0 \leq j \leq i$, and $0 \leq a \leq i-j+1$, subsequently, denote the value of their corresponding quantum bits to be 0. Moreover, the notations used in **Algorithm 4-2** below have been denoted in previous subsections. The first parameter $w$ in **Algorithm 4-2** is used to represent the maximum size of vertices among legal answers, and its value is passed from the execution of Step (1a) in **Algorithm 4-3** in next subsection. To increase the probability of success on measuring the answer from among $2^n$ choices, **Grover's algorithm** [Grover 1996] is integrated into the proposed quantum algorithm and is applied to significantly increase the amplitudes of those answers. Grover's operator is diffusion transform $G$, which is defined by matrix $G$ as follows: $G_{i,j} = (\frac{2}{2^n})$ if $i \neq j$ and $G_{i,i} = (-1 + \frac{2}{2^n})$.

Figure 4-8: The full network, **FMNO** (the abbreviation of finding the maximum number of ones), is applied to perform the first condition and the second condition in (4.6) in **Lemma 4-4**.

**Algorithm 4-2**($w$): Quantum algorithms of figuring out the number of one to legal cliques for any a graph $G = (V, E)$ with $n$ vertices and $\theta$ edges and its *complementary* graph $\overline{G} = (V, \overline{E})$ with $n$ vertices and $m = \frac{n \times (n-1)}{2} - \theta$ edges.

(1) For an initial input $|\Phi\rangle = (|1\rangle) \otimes (\otimes_{i=n}^{1} \otimes_{j=i}^{0} |z_{i,j}^{0}\rangle) \otimes (|z_{0,0}^{1}\rangle) \otimes (\otimes_{i=n-1}^{0} \otimes_{j=i}^{0} |g_{i,j}^{0}\rangle) \otimes (\otimes_{i=n-1}^{0} \otimes_{j=i}^{0} |f_{i,j}^{0}\rangle) \otimes (\otimes_{i=n-1}^{0} \otimes_{j=i}^{0} ((\otimes_{a=i-j+1}^{1} |h_{i,j,a}^{0}\rangle) \otimes (|h_{i,j,0}^{1}\rangle))) \otimes (\otimes_{k=m}^{1} |c_{k}^{0}\rangle) \otimes (|c_{0}^{1}\rangle) \otimes (\otimes_{k=m}^{1} |r_{k}^{1}\rangle) \otimes (\otimes_{b=n}^{1} |x_{b}^{0}\rangle)$, $2^n$ possible choices of $n$ bits (including all of the possible cliques) are $|\varphi_{0,0}\rangle$

$= (H) \otimes (\otimes_{i=n}^{1} \otimes_{j=i}^{0} I_{2\times 2}) \otimes (I_{2\times 2}) \otimes (\otimes_{i=n-1}^{0} \otimes_{j=i}^{0} I_{2\times 2}) \otimes (\otimes_{i=n-1}^{0} \otimes_{j=i}^{0} I_{2\times 2}) \otimes (\otimes_{i=n-1}^{0} \otimes_{j=i}^{0} ((\otimes_{a=i-j+1}^{1} I_{2\times 2}) \otimes (I_{2\times 2}))) \otimes (\otimes_{k=m}^{1} I_{2\times 2}) \otimes (I_{2\times 2}) \otimes (\otimes_{k=m}^{1} I_{2\times 2}) \otimes H^{\otimes n} |\Phi\rangle =$

$(\frac{|0\rangle - |1\rangle}{\sqrt{2}}) \otimes \frac{1}{\sqrt{2^n}} (\otimes_{i=n}^{1} \otimes_{j=i}^{0} |z_{i,j}^{0}\rangle) \otimes (|z_{0,0}^{1}\rangle) \otimes (\otimes_{i=n-1}^{0} \otimes_{j=i}^{0} |g_{i,j}^{0}\rangle) \otimes (\otimes_{i=n-1}^{0} \otimes_{j=i}^{0} |f_{i,j}^{0}\rangle) \otimes (\otimes_{i=n-1}^{0} \otimes_{j=i}^{0} ((\otimes_{a=i-j+1}^{1} |h_{i,j,a}^{0}\rangle) \otimes (|h_{i,j,0}^{1}\rangle))) \otimes (\otimes_{k=m}^{1} |c_{k}^{0}\rangle) \otimes (|c_{0}^{1}\rangle) \otimes (\otimes_{k=m}^{1} |r_{k}^{1}\rangle) \otimes (\otimes_{b=n}^{1} (|x_{b}^{0}\rangle + |x_{b}^{1}\rangle))$.

(2) $|\varphi_{1,0}\rangle = (I_{2\times 2}) \otimes (\otimes_{i=n}^{1} \otimes_{j=i}^{0} I_{2\times 2}) \otimes (I_{2\times 2}) \otimes (\otimes_{i=n-1}^{0} \otimes_{j=i}^{0} I_{2\times 2}) \otimes (\otimes_{i=n-1}^{0} \otimes_{j=i}^{0} I_{2\times 2}) \otimes (\otimes_{i=n-1}^{0} \otimes_{j=i}^{0} ((\otimes_{a=i-j+1}^{1} I_{2\times 2}) \otimes (I_{2\times 2}))) \otimes \mathbf{QEC} \otimes |\varphi_{0,0}\rangle = (\frac{|0\rangle - |1\rangle}{\sqrt{2}}) \otimes \frac{1}{\sqrt{2^n}} (\otimes_{i=n}^{1} \otimes_{j=i}^{0} |z_{i,j}^{0}\rangle) \otimes (|z_{0,0}^{1}\rangle) \otimes (\otimes_{i=n-1}^{0} \otimes_{j=i}^{0} |g_{i,j}^{0}\rangle) \otimes (\otimes_{i=n-1}^{0} \otimes_{j=i}^{0} |f_{i,j}^{0}\rangle) \otimes (\otimes_{i=n-1}^{0} \otimes_{j=i}^{0} ((\otimes_{a=i-j+1}^{1} |h_{i,j,a}^{0}\rangle) \otimes (|h_{i,j,0}^{1}\rangle))) \otimes (\otimes_{k=m}^{1} |c_{k}^{0} \oplus (c_{k-1} \bullet r_{k})\rangle) \otimes (|c_{0}^{1}\rangle) \otimes (\otimes_{k=m}^{1} |r_{k}^{1} \oplus (x_i \bullet x_j)\rangle) \otimes (\otimes_{b=n}^{1} (|x_{b}^{0}\rangle + |x_{b}^{1}\rangle))$, where **QEC** is the quantum circuit in Figure 4-7 and bits $x_i$ and $x_j$ respectively represent vertices $v_i$ and $v_j$ in the $k$th edge, $e_k = (v_i, v_j)$, in $\overline{G}$ to $1 \leq k \leq \frac{n \times (n-1)}{2} - \theta$.

(3) **For** $i = 0$ **to** $n - 1$
  (4) **For** $j = i$ **down to** 0

(4a) $\left|\varphi_{1+(\sum_{\theta_1=0}^{i-1}(\theta_1+1))+(i-j)+1, 0}\right\rangle = (I_{2\times 2}) \otimes \mathbf{FMNO} \left|\varphi_{1+(\sum_{\theta_1=0}^{i-1}(\theta_1+1))+(i-j), 0}\right\rangle = (\frac{|0\rangle - |1\rangle}{\sqrt{2}}) \otimes \frac{1}{\sqrt{2^n}}$

$(\otimes_{s=n}^{i+2} \otimes_{l=s}^{0} |z_{s,l}^{0}\rangle) \otimes (\otimes_{s=i+1}^{i+1}((\otimes_{l=s}^{j+2} |z_{s,l}\rangle) \otimes |z_{i+1, j+1} \oplus (c_m \bullet x_{i+1} \bullet z_{i,j} \bullet \overline{z}_{i+1,j+2} \bullet \overline{z}_{i+1,j+3} \bullet \cdots \bullet \overline{z}_{i+1,i+1})\rangle \otimes |z_{i+1, j} \oplus (c_m \bullet \overline{x}_{i+1} \bullet z_{i,j})\rangle \otimes (\otimes_{l=j-1}^{0} |z_{s,l}^{0}\rangle))) \otimes (\otimes_{s=i}^{1}(\otimes_{l=s}^{0} |z_{s,l}\rangle)) \otimes (|z_{0,0}^{1}\rangle) \otimes (\otimes_{d=n-1}^{i+1} \otimes_{e=d}^{0} |g_{d,e}^{0}\rangle) \otimes (\otimes_{d=i}^{i}((\otimes_{e=d}^{j+1} |g_{d,e}\rangle) \otimes |g_{i,j}^{0} \oplus (\overline{x}_{i+1} \bullet z_{i,j})\rangle \otimes (\otimes_{e=j-1}^{0} |g_{d,e}^{0}\rangle))) \otimes (\otimes_{d=i-1}^{0}(\otimes_{e=d}^{0} |g_{d,e}\rangle)) \otimes (\otimes_{d=n-1}^{i+1} \otimes_{e=d}^{0} |f_{d,e}^{0}\rangle) \otimes (\otimes_{d=i}^{i}((\otimes_{e=d}^{j+1} |f_{d,e}\rangle) \otimes |f_{i,j}^{0} \oplus (x_{i+1} \bullet h_{i,j,i-j+1})\rangle \otimes (\otimes_{e=j-1}^{0} |f_{d,e}^{0}\rangle))) \otimes (\otimes_{d=i-1}^{0}(\otimes_{e=d}^{0} |f_{d,e}\rangle)) \otimes (\otimes_{d=n-1}^{i+1} \otimes_{e=d}^{0}((\otimes_{a=d-e+1}^{1} |h_{d,e,a}^{0}\rangle) \otimes (|h_{d,e,0}^{1}\rangle))) \otimes (\otimes_{d=i}^{i}(\otimes_{e=d}^{j+1}((\otimes_{a=d-e+1}^{1} |h_{d,e,a}\rangle) \otimes (|h_{d,e,0}^{1}\rangle))) \otimes (|h_{i,j,i-j+1}^{0} \oplus (z_{i,j} \bullet h_{i,j,i-j})\rangle) \otimes (\otimes_{a=i-j}^{1}(|h_{i,j,a}^{0} \oplus (\overline{z}_{i+1,j+1+a} \bullet h_{i,j,a-1})\rangle)) \otimes (|h_{i,j,0}^{1}\rangle) \otimes (\otimes_{e=j-1}^{0}((\otimes_{a=d-e+1}^{1} |h_{d,e,a}^{0}\rangle) \otimes (|h_{d,e,0}^{1}\rangle)))) \otimes (\otimes_{d=i-1}^{0}(\otimes_{e=d}^{0}((\otimes_{a=d-e+1}^{1} |h_{d,e,a}\rangle) \otimes (|h_{d,e,0}^{1}\rangle)))) \otimes$

$(\otimes_{k=m}^{1}|c_k\rangle) \otimes (|c_0^1\rangle) \otimes (\otimes_{k=m}^{1}|r_k\rangle) \otimes (\otimes_{b=n}^{1}(|x_b^0\rangle+|x_b^1\rangle))$, where **FMNO** is the quantum circuit in Figure 4-8, and for $m \geq k \geq 1$ quantum bits $c_k$ and $r_k$ are all the results generated from Step (2).
    **End For**
**End For**

(5) $\left|\varphi_{1+\frac{n\times(n+1)}{2}+1,0}\right\rangle = (\frac{|0\rangle-|1\rangle}{\sqrt{2}} \oplus z_{n,w}) \otimes (\otimes_{i=n}^{1} \otimes_{j=i}^{0} I_{2\times 2}) \otimes (I_{2\times 2}) \otimes (\otimes_{i=n-1}^{0} \otimes_{j=i}^{0} I_{2\times 2}) \otimes$

$(\otimes_{i=n-1}^{0} \otimes_{j=i}^{0} I_{2\times 2}) \otimes (\otimes_{i=n-1}^{0} \otimes_{j=i}^{0} ((\otimes_{a=i-j+1}^{1} I_{2\times 2}) \otimes (I_{2\times 2}))) \otimes (\otimes_{k=m}^{1} I_{2\times 2}) \otimes (I_{2\times 2}) \otimes$

$(\otimes_{k=m}^{1} I_{2\times 2}) \otimes (\otimes_{b=n}^{1} I_{2\times 2}) \left|\varphi_{1+\frac{n\times(n+1)}{2},0}\right\rangle = (\frac{|0\rangle-|1\rangle}{\sqrt{2}}) \otimes \frac{1}{\sqrt{2^n}} \times (-1)^{z_{n,w}} (\otimes_{i=n}^{1} \otimes_{j=i}^{0} |z_{i,j}\rangle) \otimes$

$(|z_{0,0}^1\rangle) \otimes (\otimes_{i=n-1}^{0} \otimes_{j=i}^{0} |g_{i,j}\rangle) \otimes (\otimes_{i=n-1}^{0} \otimes_{j=i}^{0} |f_{i,j}\rangle) \otimes$

$(\otimes_{i=n-1}^{0} \otimes_{j=i}^{0} ((\otimes_{a=i-j+1}^{1} |h_{i,j,a}\rangle) \otimes (|h_{i,j,0}^1\rangle))) \otimes (\otimes_{k=m}^{1} |c_k\rangle) \otimes (|c_0^1\rangle) \otimes (\otimes_{k=m}^{1} |r_k\rangle) \otimes$

$(\otimes_{b=n}^{1}(|x_b^0\rangle+|x_b^1\rangle))$.

(6) Since quantum operations are reversible by nature, the auxiliary quantum bits can be restored back to their initial states by reversing all these operations finished by Step (4a) through Step (2).

(7) Apply Grover's operator in **Grover's algorithm** to the quantum state vector generated by Step (6).

(8) At most repeat to execute from Step (2) to Step (7) of $2^{\frac{n}{2}}$ times.

(9) The answer is obtained with a successful probability of at least $\frac{1}{2}$ after a measurement is finished and return the result to **Algorithm 4-3**.
**EndAlgorithm**

**Lemma 4-6**: **Algorithm 4-2** is used to calculate the number of ones (vertices) to legal cliques for any graph $G = (V, E)$ with $n$ vertices and $\theta$ edges and its *complementary* graph $\overline{G} = (V, \overline{E})$ with $n$ vertices and $m = \frac{n\times(n-1)}{2} - \theta$ edges.

**Proof**:

    Since there are $2^n$ possible choices (including all possible cliques) for the clique problem of any graph $G$ with $\theta$ edges and $n$ vertices, a quantum register of $n$ bits $(\otimes_{b=n}^{1}|x_b\rangle)$ is used to stand for $2^n$ choices with initial state $(\otimes_{b=n}^{1}|x_b^0\rangle)$. The clique problem of any graph $G$ with $\theta$ edges and $n$ vertices requires finding a maximum-sized clique in $G$, so those auxiliary quantum registers are needed. From the execution of Step (1), an initial vector $|\Phi\rangle$

$= (|1\rangle) \otimes (\otimes_{i=n}^{1} \otimes_{j=i}^{0} |z_{i,j}^0\rangle) \otimes (|z_{0,0}^1\rangle) \otimes (\otimes_{i=n-1}^{0} \otimes_{j=i}^{0} |g_{i,j}^0\rangle) \otimes (\otimes_{i=n-1}^{0} \otimes_{j=i}^{0} |f_{i,j}^0\rangle) \otimes$

$(\otimes_{i=n-1}^{0} \otimes_{j=i}^{0} ((\otimes_{a=i-j+1}^{1} |h_{i,j,a}^0\rangle) \otimes (|h_{i,j,0}^1\rangle))) \otimes (\otimes_{k=m}^{1} |c_k^0\rangle) \otimes (|c_0^1\rangle) \otimes (\otimes_{k=m}^{1} |r_k^1\rangle) \otimes$

$(\otimes_{b=n}^{1}|x_b^0\rangle)$ starts the quantum computation of the clique problem. $H^{\otimes n}$ that represent the joined $n$-qubit Hadamard gate is used to the part of choices of the initial vector $|\Phi\rangle$, and then the resulting state vector becomes $|\varphi_{0,0}\rangle$ with $2^n$ choices. This indicates that the function of Step (1) in **Algorithm 4-1** can be implemented by Step (1) in **Algorithm 4-2**.

Next Step (2) in **Algorithm 4-2** works as the unitary operator **QEC** which is the quantum circuit in Figure 4-7. On the execution of Step (2) in **Algorithm 4-2**, it is employed to find choices among $2^n$ possible choices that satisfy each formula of the form $\overline{x_i \wedge x_j}$. After Step (2) in **Algorithm 4-2** is executed, the resulting state vector $|\varphi_{1,0}\rangle$ is obtained containing those legal choices with $|c_m^1\rangle$ that do not contain any edge in $\overline{G}$ and those illegal choices with $|c_m^0\rangle$ that dissatisfy the condition. This indicates that the function performed by Steps (2a) through (2d) in **Algorithm 4-1** can be implemented by Step (2) in **Algorithm 4-2**.

Next, Step (4a) is embedded in the first loop in **Algorithm 4-2** and works as the unitary operator **FMNO,** which is the quantum circuit in Figure 4-8. The step is applied to calculate the number of ones (the number of vertices) among the legal choices. After repeating to execute Step (4a), the resulting state vector $\left|\varphi_{1+\frac{n\times(n+1)}{2},0}\right\rangle$ is obtained in which the number of vertices in each legal clique is calculated. This implies that Steps (4a) through (4b) in **Algorithm 4-1** can be implemented by Step (4a) in **Algorithm 4-2**.

Next, one **CNOT** gate, $(\frac{|0\rangle-|1\rangle}{\sqrt{2}} \oplus z_{n,w})$, in the execution of Step (5) in **Algorithm 4-2** is used to perform the oracle work (in the language of **Grover's algorithm**), that is, the target state labeling preceding Grover's searching step. The resulting state vector $\left|\varphi_{1+\frac{n\times(n+1)}{2}+1,0}\right\rangle$ contains the part of the answer with phase $-1$ and the other part with phase $+1$. Since quantum operations are reversible by nature, the auxiliary quantum bits can be restored to their initial states by simply applying the reversible operation, and then they can be repeated for safe use. Therefore, the execution of Step (6) in **Algorithm 4-2** is used to reverse all those operations completed by Steps (4a) and (2). Next, on the execution of Step (7) in **Algorithm 4-2**, it applies Grover's operator in **Grover's algorithm** to perform the task of increasing the probability of success in measuring the answer. From Step (8) in **Algorithm 4-2**, after repeating to execute Steps (2) through (7) for $2^{\frac{n}{2}}$ times, a maximum successful probability is generated. Next, from the execution of Step (9) in **Algorithm 4-2**, a measurement is used to obtain the answer(s) and the answer(s) is/are returned to **Algorithm 4-3**. Therefore, it is at once inferred that **Algorithm 4-2** can be used to calculate the number of ones (vertices) in legal cliques for any graph $G = (V, E)$ with $n$ vertices and $\theta$ edges and its *complementary* graph $\overline{G} = (V, \overline{E})$ with $n$ vertices and $m = \frac{n\times(n-1)}{2} - \theta$ edges on a quantum computer. ∎

4.11. QUANTUM ALGORITHMS FOR SOLVING AN INSTANCE OF THE CLIQUE PROBLEM OF ANY GRAPH WITH θ EDGES AND *N* VERTICES

The following quantum algorithm is applied to solve an instance of the clique problem of any graph $G$ with $\theta$ edges and $n$ vertices. The notations used in **Algorithm 4-3** below have been denoted in previous subsections.

**Algorithm 4-3**: Quantum algorithms for solving an instance of the clique problem for any a graph $G = (V, E)$ with $n$ vertices and $\theta$ edges and its *complementary* graph $\overline{G} = (V, \overline{E})$ with $n$ vertices and $m = \frac{n\times(n-1)}{2} - \theta$ edges.

(1) **For** $w = n$ **to** 1
    (1a) Call **Algorithm 4-2**($w$).

(1b) **If** the answer is obtained from the *w*th execution of Step (1a) **then**
    (1c) Terminate **Algorithm 4-3**.
  **End If**
**End For**
**End Algorithm**

Consider the graph $G_1 = (V, E)$ in Figure 4-1. Its vertices are, respectively, $v_1$ and $v_2$ and its edge is $(v_1, v_2)$. This is to say that $V = \{v_1, v_2\}$ and $E = \{(v_1, v_2)\}$. For the graph $G_1$ in Figure 4-1, its *complementary* graph $\overline{G}_1 = (V, \overline{E})$, where $V = \{v_1, v_2\}$ and $\overline{E}$ is an empty set that does not include any an edge. **Algorithm 4-3** is used to search the answer of the clique problem for the graph $G_1$ in Figure 4-1. Because in the graph $G_1$ in Figure 4-1 the number of vertices is two and the number of edges is one, the value of $n$ is equal to two and the value of $m$ is equal to zero ($\frac{2\times(2-1)}{2} - 1$), where the value of $m$ is applied to represent the number of edges for its *complementary* graph, $\overline{G}_1$. Step (1) is the only loop in **Algorithm 4-3** and is employed to figure out the answer. Therefore, on the first execution of Step (1a), it calls **Algorithm 4-2**. The value of the first parameter $w$ in the first execution of Step (1a) in **Algorithm 4-3** is equal to two. From Step (1) of **Algorithm 4-2**, for an initial input $|\Phi\rangle = (|1\rangle) \otimes (|z_{2,2}{}^0\rangle) \otimes |z_{2,1}{}^0\rangle \otimes |z_{2,0}{}^0\rangle \otimes |z_{1,1}{}^0\rangle \otimes |z_{1,0}{}^0\rangle \otimes |z_{0,0}{}^1\rangle \otimes (|g_{1,1}{}^0\rangle) \otimes (|g_{1,0}{}^0\rangle) \otimes (|g_{0,0}{}^0\rangle) \otimes (|f_{1,1}{}^0\rangle) \otimes (|f_{1,0}{}^0\rangle) \otimes (|f_{0,0}{}^0\rangle) \otimes (|h_{1,1,1}{}^0\rangle \otimes |h_{1,1,0}{}^1\rangle) \otimes (|h_{1,0,2}{}^0\rangle \otimes |h_{1,0,1}{}^0\rangle \otimes |h_{1,0,0}{}^1\rangle) \otimes (|h_{0,0,1}{}^0\rangle \otimes |h_{0,0,0}{}^1\rangle) \otimes (|c_0^1\rangle) \otimes (|x_2{}^0\rangle \otimes |x_1{}^0\rangle)$, $2^2$ possible choices are $|\varphi_{0,0}\rangle = (\frac{|0\rangle - |1\rangle}{\sqrt{2}}) \otimes (\frac{1}{2}(\otimes_{i=2}^1 \otimes_{j=i}^0 |z_{i,j}{}^0\rangle) \otimes (|z_{0,0}{}^1\rangle) \otimes (\otimes_{i=1}^0 \otimes_{j=i}^0 |g_{i,j}{}^0\rangle) \otimes (\otimes_{i=1}^0 \otimes_{j=i}^0 |f_{i,j}{}^0\rangle) \otimes (\otimes_{i=1}^0 \otimes_{j=i}^0 ((\otimes_{a=i-j+1}^1 |h_{i,j,a}{}^0\rangle) \otimes (|h_{i,j,0}{}^1\rangle))) \otimes (|c_0^1\rangle) \otimes (|x_2{}^0 x_1{}^0\rangle + |x_2{}^0 x_1{}^1\rangle + |x_2{}^1 x_1{}^0\rangle + |x_2{}^1 x_1{}^1\rangle))$. Since there is no edge in $\overline{G}_1$ from the *first* executi-

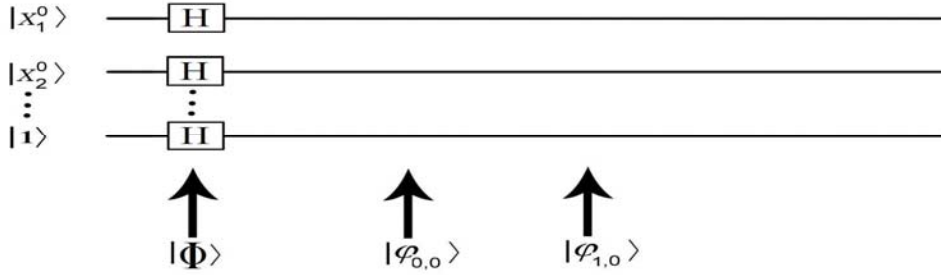

Figure 4-9: The quantum circuit for the first execution of Step (1) and Step (2) in **Algorithm 4-2**

on of Step (2) in **Algorithm 4-2**, the corresponding quantum circuit is shown in Figure 4-9. From this figure, the following is obtained: $|\varphi_{1,0}\rangle = |\varphi_{0,0}\rangle$. This is to say that $2^2$ possible choices with $|c_0^1\rangle$ in $|\varphi_{1,0}\rangle$ are all legal cliques.

Next, from the *first* execution of Step (4a) in **Algorithm 4-2**, the corresponding quantum circuit is shown in Figure 4-10. From this figure, the following is obtained: $|\varphi_{2,0}\rangle = (\frac{|0\rangle - |1\rangle}{\sqrt{2}}) \otimes (\frac{1}{2}(\otimes_{s=2}^1 \otimes_{l=s}^0 |z_{s,l}{}^0\rangle) \otimes ((|z_{1,1}{}^0\rangle) \otimes (|z_{1,0}{}^1\rangle) \otimes (|z_{0,0}{}^1\rangle) \otimes (|g_{1,1}{}^0\rangle) \otimes (|g_{1,0}{}^0\rangle) \otimes (|g_{0,0}{}^1\rangle) \otimes (|f_{1,1}{}^0\rangle) \otimes (|f_{1,0}{}^0\rangle) \otimes (|f_{0,0}{}^0\rangle) \otimes$

$(|h_{1,1,1}{}^{0}\rangle \otimes |h_{1,1,0}{}^{1}\rangle) \otimes (|h_{1,0,2}{}^{0}\rangle \otimes |h_{1,0,1}{}^{0}\rangle \otimes |h_{1,0,0}{}^{1}\rangle) \otimes (|h_{0,0,1}{}^{1}\rangle \otimes |h_{0,0,0}{}^{1}\rangle) \otimes (|c_{0}{}^{1}\rangle) \otimes (|x_{2}{}^{0}x_{1}{}^{0}\rangle)$
$+ (|z_{1,1}{}^{1}\rangle) \otimes (|z_{1,0}{}^{0}\rangle) \otimes (|z_{0,0}{}^{1}\rangle) \otimes (|g_{1,1}{}^{0}\rangle) \otimes (|g_{1,0}{}^{0}\rangle) \otimes (|g_{0,0}{}^{0}\rangle) \otimes (|f_{1,1}{}^{0}\rangle) \otimes (|f_{1,0}{}^{0}\rangle) \otimes (|f_{0,0}{}^{1}\rangle) \otimes$
$|h_{1,1,1}{}^{0}\rangle \otimes |h_{1,1,0}{}^{1}\rangle) \otimes (|h_{1,0,2}{}^{0}\rangle \otimes |h_{1,0,1}{}^{0}\rangle \otimes |h_{1,0,0}{}^{1}\rangle) \otimes (|h_{0,0,1}{}^{1}\rangle \otimes |h_{0,0,0}{}^{1}\rangle) \otimes (|c_{0}{}^{1}\rangle) \otimes (|x_{2}{}^{0}x_{1}{}^{1}\rangle)$

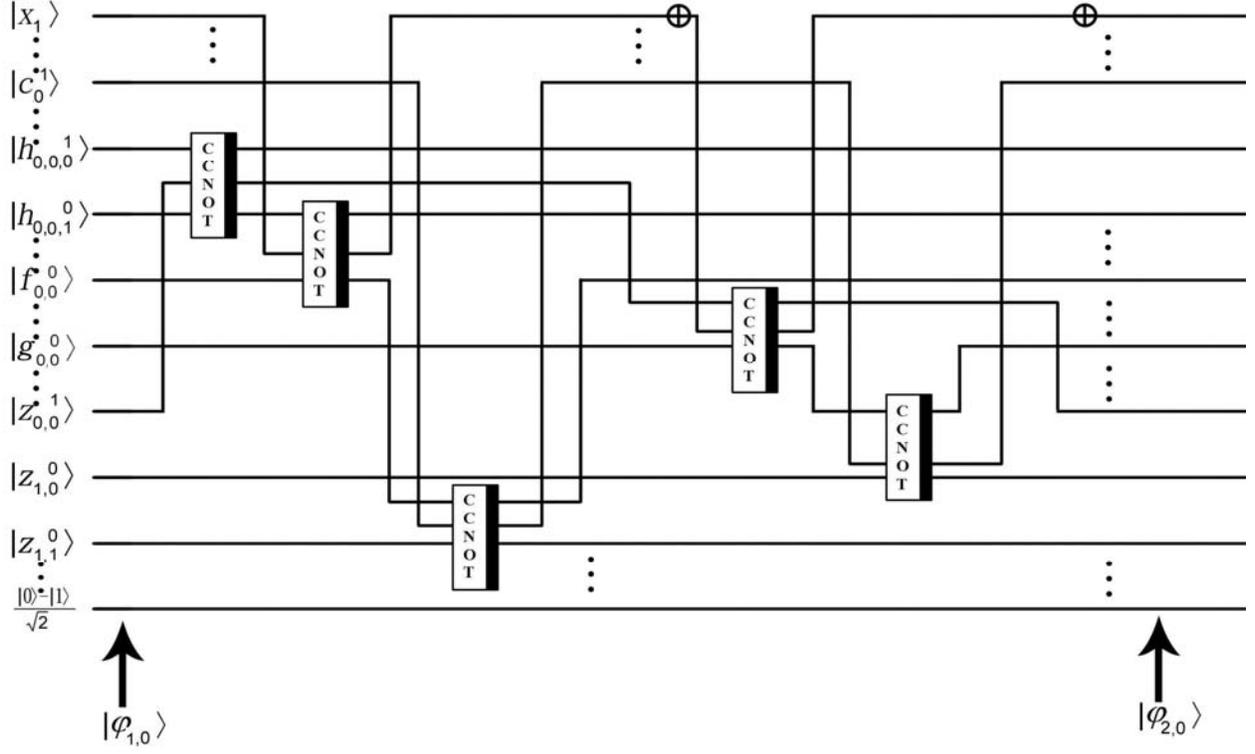

Figure 4-10: The quantum circuit for the first execution of Step (4a) in **Algorithm 4-2**

$+ (|z_{1,1}{}^{0}\rangle) \otimes (|z_{1,0}{}^{1}\rangle) \otimes (|z_{0,0}{}^{1}\rangle) \otimes (|g_{1,1}{}^{0}\rangle) \otimes (|g_{1,0}{}^{0}\rangle) \otimes (|g_{0,0}{}^{1}\rangle) \otimes (|f_{1,1}{}^{0}\rangle) \otimes (|f_{1,0}{}^{0}\rangle) \otimes (|f_{0,0}{}^{0}\rangle) \otimes$
$(|h_{1,1,1}{}^{0}\rangle \otimes |h_{1,1,0}{}^{1}\rangle) \otimes (|h_{1,0,2}{}^{0}\rangle \otimes |h_{1,0,1}{}^{0}\rangle \otimes |h_{1,0,0}{}^{1}\rangle) \otimes (|h_{0,0,1}{}^{1}\rangle \otimes |h_{0,0,0}{}^{1}\rangle) \otimes (|c_{0}{}^{1}\rangle) \otimes (|x_{2}{}^{1}x_{1}{}^{0}\rangle)$
$+ (|z_{1,1}{}^{1}\rangle) \otimes (|z_{1,0}{}^{0}\rangle) \otimes (|z_{0,0}{}^{1}\rangle) \otimes (|g_{1,1}{}^{0}\rangle) \otimes (|g_{1,0}{}^{0}\rangle) \otimes (|g_{0,0}{}^{0}\rangle) \otimes (|f_{1,1}{}^{0}\rangle) \otimes (|f_{1,0}{}^{0}\rangle) \otimes (|f_{0,0}{}^{1}\rangle) \otimes$
$(|h_{1,1,1}{}^{0}\rangle \otimes |h_{1,1,0}{}^{1}\rangle) \otimes (|h_{1,0,2}{}^{0}\rangle \otimes |h_{1,0,1}{}^{0}\rangle \otimes |h_{1,0,0}{}^{1}\rangle) \otimes (|h_{0,0,1}{}^{1}\rangle \otimes |h_{0,0,0}{}^{1}\rangle) \otimes (|c_{0}{}^{1}\rangle) \otimes$
$(|x_{2}{}^{1}x_{1}{}^{1}\rangle))$.

Next, from the *second* execution of Step (4a) in **Algorithm 4-2**, the corresponding quantum circuit is shown in Figure 4-11. From this figure, the following is obtained: $|\varphi_{3,0}\rangle = (\frac{|0\rangle - |1\rangle}{\sqrt{2}}) \otimes (\frac{1}{2}(((|z_{2,2}{}^{0}\rangle) \otimes (|z_{2,1}{}^{0}\rangle) \otimes$
$(|z_{2,0}{}^{0}\rangle) \otimes (|z_{1,1}{}^{0}\rangle) \otimes (|z_{1,0}{}^{1}\rangle) \otimes (|z_{0,0}{}^{1}\rangle) \otimes (|g_{1,1}{}^{0}\rangle) \otimes (|g_{1,0}{}^{0}\rangle) \otimes (|g_{0,0}{}^{1}\rangle) \otimes (|f_{1,1}{}^{0}\rangle) \otimes (|f_{1,0}{}^{0}\rangle) \otimes$
$(|f_{0,0}{}^{0}\rangle) \otimes (|h_{1,1,1}{}^{0}\rangle \otimes |h_{1,1,0}{}^{1}\rangle) \otimes (|h_{1,0,2}{}^{0}\rangle \otimes |h_{1,0,1}{}^{0}\rangle \otimes |h_{1,0,0}{}^{1}\rangle) \otimes (|h_{0,0,1}{}^{1}\rangle \otimes |h_{0,0,0}{}^{1}\rangle) \otimes (|c_{0}{}^{1}\rangle)$

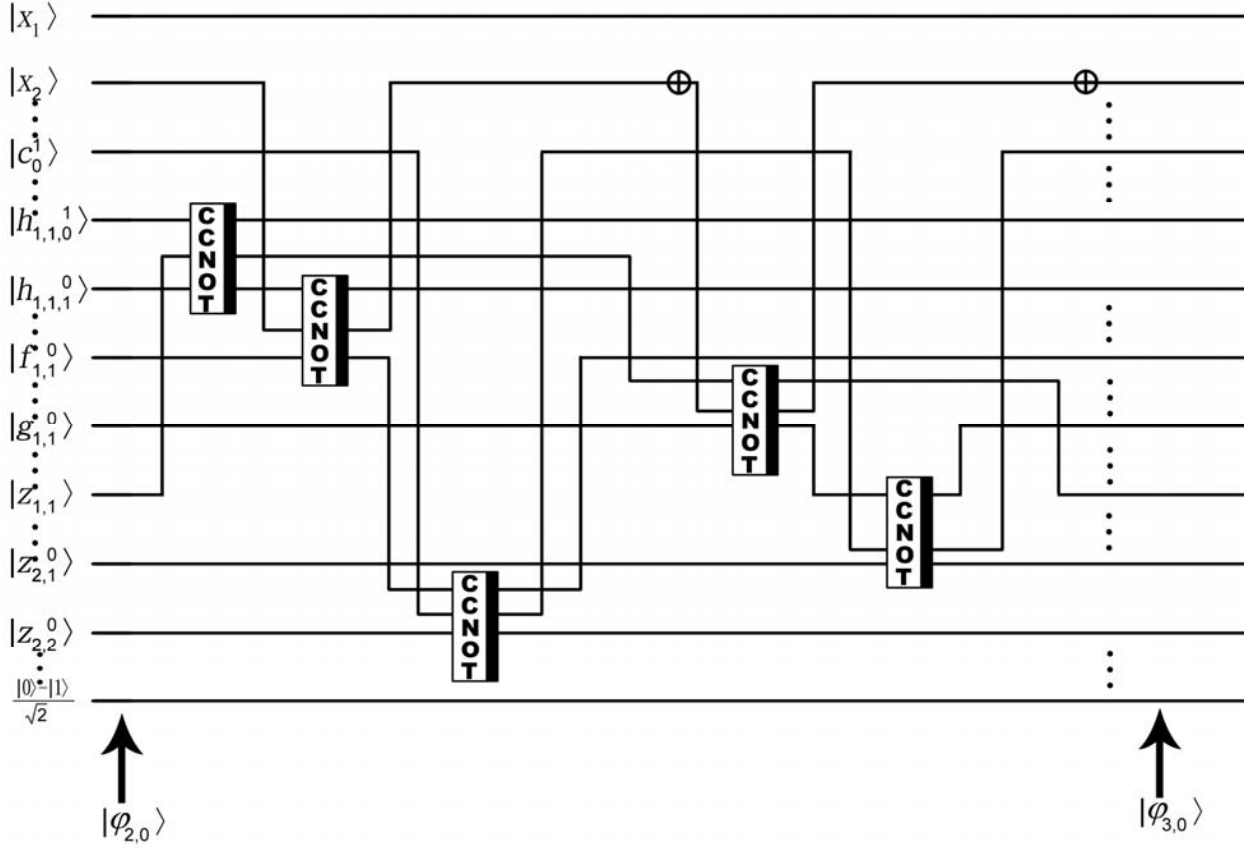

Figure 4-11: The quantum circuit for the second execution of Step (4a) in **Algorithm 4-2**

$\otimes (|x_2^0 x_1^0\rangle) + (|z_{2,2}^0\rangle) \otimes (|z_{2,1}^1\rangle) \otimes (|z_{2,0}^0\rangle) \otimes (|z_{1,1}^1\rangle) \otimes (|z_{1,0}^0\rangle) \otimes (|z_{0,0}^1\rangle) \otimes (|g_{1,1}^1\rangle) \otimes (|g_{1,0}^0\rangle) \otimes$
$(|g_{0,0}^0\rangle) \otimes (|f_{1,1}^0\rangle) \otimes (|f_{1,0}^0\rangle) \otimes (|f_{0,0}^1\rangle) \otimes (|h_{1,1,1}^1\rangle \otimes |h_{1,1,0}^1\rangle) \otimes (|h_{1,0,2}^0\rangle \otimes |h_{1,0,1}^0\rangle \otimes |h_{1,0,0}^1\rangle)$
$\otimes (|h_{0,0,1}^1\rangle \otimes |h_{0,0,0}^1\rangle) \otimes (|c_0^1\rangle) \otimes (|x_2^0 x_1^1\rangle) + (|z_{2,2}^0\rangle) \otimes (|z_{2,1}^1\rangle) \otimes (|z_{2,0}^0\rangle) \otimes (|z_{1,1}^0\rangle) \otimes (|z_{1,0}^1\rangle) \otimes$
$(|z_{0,0}^1\rangle) \otimes (|g_{1,1}^0\rangle) \otimes (|g_{1,0}^0\rangle) \otimes (|g_{0,0}^1\rangle) \otimes (|f_{1,1}^0\rangle) \otimes (|f_{1,0}^0\rangle) \otimes (|f_{0,0}^0\rangle) \otimes (|h_{1,1,1}^0\rangle \otimes |h_{1,1,0}^1\rangle) \otimes$
$(|h_{1,0,2}^0\rangle \otimes |h_{1,0,1}^0\rangle \otimes |h_{1,0,0}^1\rangle) \otimes (|h_{0,0,1}^1\rangle \otimes |h_{0,0,0}^1\rangle) \otimes (|c_0^1\rangle) \otimes (|x_2^0 x_1^0\rangle) + (|z_{2,2}^1\rangle) \otimes (|z_{2,1}^0\rangle) \otimes$
$(|z_{2,0}^0\rangle) \otimes (|z_{1,1}^1\rangle) \otimes (|z_{1,0}^0\rangle) \otimes (|z_{0,0}^1\rangle) \otimes (|g_{1,1}^0\rangle) \otimes (|g_{1,0}^0\rangle) \otimes (|g_{0,0}^0\rangle) \otimes (|f_{1,1}^1\rangle) \otimes (|f_{1,0}^0\rangle) \otimes$
$(|f_{0,0}^1\rangle) \otimes (|h_{1,1,1}^1\rangle \otimes |h_{1,1,0}^1\rangle) \otimes (|h_{1,0,2}^0\rangle \otimes |h_{1,0,1}^0\rangle \otimes |h_{1,0,0}^1\rangle) \otimes (|h_{0,0,1}^1\rangle \otimes |h_{0,0,0}^1\rangle) \otimes (|c_0^1\rangle) \otimes$
$(|x_2^1 x_1^1\rangle))$.

Next, from the *third* execution of Step (4a) in **Algorithm 4-2**, the corresponding quantum circuit is shown in Figure 4-12. From this figure, the following is obtained: $|\varphi_{4,0}\rangle = (\frac{|0\rangle - |1\rangle}{\sqrt{2}}) \otimes (\frac{1}{2}((|z_{2,2}^0\rangle) \otimes (|z_{2,1}^0\rangle) \otimes$
$(|z_{2,0}^1\rangle) \otimes (|z_{1,1}^0\rangle) \otimes (|z_{1,0}^1\rangle) \otimes (|z_{0,0}^1\rangle) \otimes (|g_{1,1}^0\rangle) \otimes (|g_{1,0}^1\rangle) \otimes (|g_{0,0}^1\rangle) \otimes (|f_{1,1}^0\rangle) \otimes (|f_{1,0}^0\rangle) \otimes$

$(|f_{0,0}{}^0\rangle) \otimes (|h_{1,1,1}{}^0\rangle \otimes |h_{1,1,0}{}^1\rangle) \otimes (|h_{1,0,2}{}^1\rangle \otimes |h_{1,0,1}{}^1\rangle \otimes |h_{1,0,0}{}^1\rangle) \otimes (|h_{0,0,1}{}^1\rangle \otimes |h_{0,0,0}{}^1\rangle) \otimes (|c_0{}^1\rangle) \otimes (|x_2{}^0 x_1{}^0\rangle) + (|z_{2,2}{}^0\rangle) \otimes (|z_{2,1}{}^1\rangle) \otimes (|z_{2,0}{}^0\rangle) \otimes (|z_{1,1}{}^1\rangle) \otimes (|z_{1,0}{}^0\rangle) \otimes (|z_{0,0}{}^1\rangle) \otimes (|g_{1,1}{}^1\rangle) \otimes (|g_{1,0}{}^0\rangle) \otimes (|g_{0,0}{}^0\rangle) \otimes (|f_{1,1}{}^0\rangle) \otimes (|f_{1,0}{}^0\rangle) \otimes (|f_{0,0}{}^1\rangle) \otimes (|h_{1,1,1}{}^1\rangle \otimes |h_{1,1,0}{}^1\rangle) \otimes (|h_{1,0,2}{}^0\rangle \otimes |h_{1,0,1}{}^1\rangle \otimes |h_{1,0,0}{}^1\rangle) \otimes (|h_{0,0,1}{}^1\rangle \otimes |h_{0,0,0}{}^1\rangle) \otimes (|c_0{}^1\rangle) \otimes (|x_2{}^0 x_1{}^1\rangle) + (|z_{2,2}{}^0\rangle) \otimes (|z_{2,1}{}^1\rangle) \otimes (|z_{2,0}{}^0\rangle) \otimes (|z_{1,1}{}^0\rangle) \otimes (|z_{1,0}{}^1\rangle) \otimes (|z_{0,0}{}^1\rangle) \otimes (|g_{1,1}{}^0\rangle) \otimes (|g_{1,0}{}^0\rangle) \otimes (|g_{0,0}{}^1\rangle) \otimes (|f_{1,1}{}^0\rangle) \otimes (|f_{1,0}{}^1\rangle) \otimes (|f_{0,0}{}^0\rangle) \otimes (|h_{1,1,1}{}^0\rangle \otimes |h_{1,1,0}{}^1\rangle) \otimes (|h_{1,0,2}{}^1\rangle \otimes |h_{1,0,1}{}^1\rangle \otimes |h_{1,0,0}{}^1\rangle) \otimes (|h_{0,0,1}{}^1\rangle \otimes |h_{0,0,0}{}^1\rangle) \otimes (|c_0{}^1\rangle) \otimes (|x_2{}^1 x_1{}^0\rangle) + (|z_{2,2}{}^1\rangle) \otimes (|z_{2,1}{}^0\rangle) \otimes (|z_{2,0}{}^0\rangle) \otimes (|z_{1,1}{}^1\rangle) \otimes (|z_{1,0}{}^0\rangle) \otimes (|z_{0,0}{}^1\rangle) \otimes (|g_{1,1}{}^0\rangle) \otimes (|g_{1,0}{}^0\rangle) \otimes (|g_{0,0}{}^0\rangle) \otimes (|f_{1,1}{}^1\rangle) \otimes (|f_{1,0}{}^0\rangle) \otimes (|f_{0,0}{}^1\rangle) \otimes (|h_{1,1,1}{}^1\rangle \otimes |h_{1,1,0}{}^1\rangle) \otimes (|h_{1,0,2}{}^0\rangle \otimes |h_{1,0,1}{}^0\rangle \otimes |h_{1,0,0}{}^1\rangle) \otimes (|h_{0,0,1}{}^1\rangle \otimes |h_{0,0,0}{}^1\rangle) \otimes (|c_0{}^1\rangle) \otimes (|x_2{}^1 x_1{}^1\rangle))$.

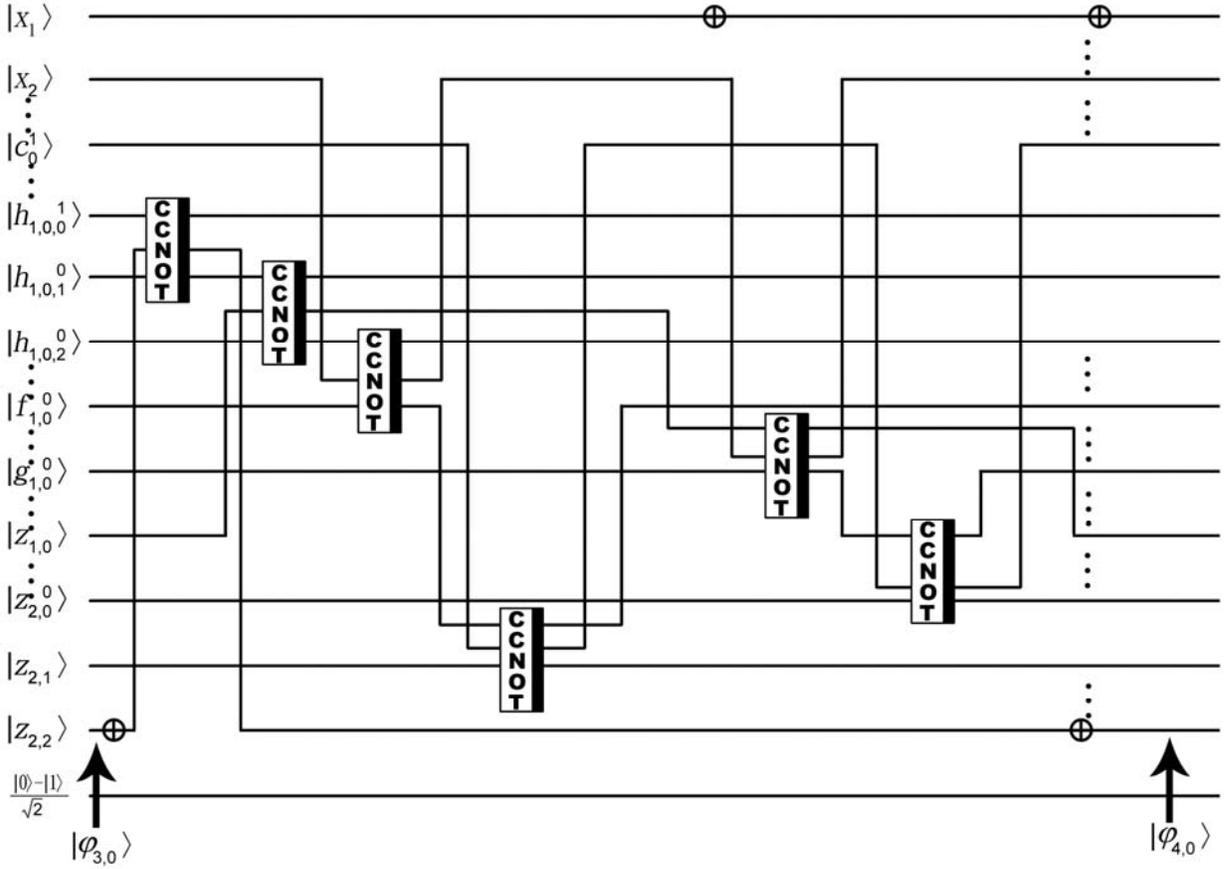

Figure 4-12: The quantum circuit for the third execution of Step (4a) in **Algorithm 4-2**

Next, after the first execution of Step (5) in **Algorithm 4-2**, wherein a **CNOT** gate $(\frac{|0\rangle - |1\rangle}{\sqrt{2}} \oplus z_{2,2})$ is

completed, $|\varphi_{5,0}\rangle$ is obtained. In $|\varphi_{5,0}\rangle$ for $|x_2^1 x_1^1\rangle$ its amplitude is $-\frac{1}{2}$, and for three other choices (states), their amplitudes are all $\frac{1}{2}$. This is to say that the first oracle work is completed from the execution of Steps (2) through (5). Next, from the first execution of Step (6) in **Algorithm 4-2**, those auxiliary quantum bits are all restored to their initial states. Therefore, the following is obtained: $|\varphi_{6,0}\rangle = (\frac{|0\rangle - |1\rangle}{\sqrt{2}}) \otimes ((|z_{2,2}^0\rangle \otimes |z_{2,1}^0\rangle \otimes |z_{2,0}^0\rangle \otimes |z_{1,1}^0\rangle \otimes |z_{1,0}^0\rangle \otimes |z_{0,0}^1\rangle) \otimes (|g_{1,1}^0\rangle) \otimes (|g_{1,0}^0\rangle) \otimes (|g_{0,0}^0\rangle) \otimes (|f_{1,1}^0\rangle) \otimes (|f_{1,0}^0\rangle) \otimes (|f_{0,0}^0\rangle) \otimes (|h_{1,1,1}^0\rangle \otimes |h_{1,1,0}^1\rangle) \otimes (|h_{1,0,2}^0\rangle \otimes |h_{1,0,1}^0\rangle \otimes |h_{1,0,0}^1\rangle) \otimes (|h_{0,0,1}^0\rangle \otimes |h_{0,0,0}^1\rangle) \otimes (|c_0^1\rangle) \otimes (\frac{1}{2} |x_2^0 x_1^0\rangle + \frac{1}{2} |x_2^0 x_1^1\rangle + \frac{1}{2} |x_2^1 x_1^0\rangle + (-\frac{1}{2}) |x_2^1 x_1^1\rangle))$.

Next, in the first execution of Step (7) in **Algorithm 4-2**, Grover's operator in **Grover's algorithm** is used to increase the amplitude in the quantum state vector $|\varphi_{6,0}\rangle$ generated from the first execution of Step (6) in **Algorithm 4-2**. Grover's operator allows the inversion with respect to the average state. The subspace spanned by $\{|x_2\rangle, |x_1\rangle\}$ has the Grover's operator that is a diffusion transform which is defined by a matrix $G$ (a $4 \times 4$ matrix) as follows: $G_{i,j} = (\frac{2}{2^2})$ if $i \neq j$ and $G_{i,i} = (-1 + \frac{2}{2^2})$. After one oracle work and one **Grover's** operator are performed, the following is obtained: $|\varphi_{7,0}\rangle = (\frac{|0\rangle - |1\rangle}{\sqrt{2}}) \otimes ((|z_{2,2}^0\rangle \otimes |z_{2,1}^0\rangle \otimes |z_{2,0}^0\rangle \otimes |z_{1,1}^0\rangle \otimes |z_{1,0}^0\rangle \otimes |z_{0,0}^1\rangle) \otimes (|g_{1,1}^0\rangle) \otimes (|g_{1,0}^0\rangle) \otimes (|g_{0,0}^0\rangle) \otimes (|f_{1,1}^0\rangle) \otimes (|f_{1,0}^0\rangle) \otimes (|f_{0,0}^0\rangle) \otimes (|h_{1,1,1}^0\rangle \otimes |h_{1,1,0}^1\rangle) \otimes (|h_{1,0,2}^0\rangle \otimes |h_{1,0,1}^0\rangle \otimes |h_{1,0,0}^1\rangle) \otimes (|h_{0,0,1}^0\rangle \otimes |h_{0,0,0}^1\rangle) \otimes (|c_0^1\rangle) \otimes (|x_2^1 x_1^1\rangle))$. Therefore, from Step (8) in **Algorithm 4-2**, we do not need to repeat to execute Step (2) to Step (7).

Next, from the first execution of Step (9) in **Algorithm 4-2** and after a measurement on $|\varphi_{7,0}\rangle$ is performed, the answer with the successful probability of one is $(|z_{2,2}^0\rangle \otimes |z_{2,1}^0\rangle \otimes |z_{2,0}^0\rangle \otimes |z_{1,1}^0\rangle \otimes |z_{1,0}^0\rangle \otimes |z_{0,0}^1\rangle) \otimes (|g_{1,1}^0\rangle) \otimes (|g_{1,0}^0\rangle) \otimes (|g_{0,0}^0\rangle) \otimes (|f_{1,1}^0\rangle) \otimes (|f_{1,0}^0\rangle) \otimes (|f_{0,0}^0\rangle) \otimes (|h_{1,1,1}^0\rangle \otimes |h_{1,1,0}^1\rangle) \otimes (|h_{1,0,2}^0\rangle \otimes |h_{1,0,1}^0\rangle \otimes |h_{1,0,0}^1\rangle) \otimes (|h_{0,0,1}^0\rangle \otimes |h_{0,0,0}^1\rangle) \otimes (|c_0^1\rangle) \otimes (|x_2^1 x_1^1\rangle)$.

The result of **Algorithm 4-2** is returned to Step (1a) in **Algorithm 4-3**. Next, from the first execution of Step (1b) in **Algorithm 4-3**, this implies that a maximum-sized clique for the clique problem in Figure 4-1 is $\{v_1, v_2\}$. Therefore, **Algorithm 4-3** is terminated from the first execution of Step (1c) in **Algorithm 4-3**. The following **Lemma** is used to prove the correction of **Algorithm 4-3**.

**Lemma 4-7**: **Algorithm 4-3** is the quantum implementation of **Algorithm 4-1** (the DNA-based algorithm) which is equivalent to the oracle work (in the language of **Grover's Algorithm**), that is, the target state labeling preceding Grover's searching step, and is used to solve an instance of the clique problem for any a graph $G = (V, E)$ with $n$ vertices and $\theta$ edges and its *complementary* graph $\overline{G} = (V, \overline{E})$ with $n$ vertices and $m = \frac{n \times (n-1)}{2} - \theta$ edges.

**Proof**:

In each execution of Step (1a) in **Algorithm 4-3**, **Algorithm 4-2** is used to perform the two main tasks. The first main task is to compute the number of vertices in each legal clique. This is to say that from **Lemma 4-6** the oracle work in the language of **Grover's algorithm**, that is, the target state labeling preceding Grover's searching step, can be implemented by **Algorithm 4-2**. The second main task is to call for **Grover's algorithm** that increases the probability of success in measuring the answer from the maximum-sized cliques. Next, in each execution of Step (1b) in **Algorithm 4-3**, if the answer is found from the $w$th execution of Step (1a) in **Algorithm 4-3**, then the $w$th execution of Step (1c) in **Algorithm 4-3** is applied to terminate **Algorithm 4-3**. Otherwise, Steps (1a) through (1c) are executed until the answer to solving an instance of the clique problem is found for any a graph $G = (V, E)$ with $n$ vertices and $\theta$ edges and its *complementary* graph $\overline{G} = (V, \overline{E})$ with $n$ vertices and $m = \dfrac{n \times (n-1)}{2} - \theta$ edges.

∎

## 5. COMPLEXITY ASSESSMENT

The following **lemmas** are employed to demonstrate the time complexity and the space complexity of **Algorithm 4-3** for solving an instance of the clique problem of any graph $G = (V, E)$ with $n$ vertices and $\theta$ edges and its *complementary* graph $\overline{G} = (V, \overline{E})$ with $n$ vertices and $m = \dfrac{n \times (n-1)}{2} - \theta$ edges.

**Lemma 5-1**: The best case for time complexity of solving an instance of the clique problem for any graph $G = (V, E)$ with $n$ vertices and $\theta$ edges and its *complementary* graph $\overline{G} = (V, \overline{E})$ with $n$ vertices and $m = \dfrac{n \times (n-1)}{2} - \theta$ edges is $O(n + 1)$ **Hadamard** gates, $O(2^{\frac{n}{2}} \times (\dfrac{2 \times n^3 + 6 \times n^2 + 4 \times n}{3}))$ **NOT** gates, $O(2^{\frac{n}{2}})$ **CNOT** gates, $O(2^{\frac{n}{2}} \times (4 \times m + (\dfrac{n^3 + 15 \times n^2 + 14 \times n}{3})))$ **CCNOT** gates, $O(2^{\frac{n}{2}})$ Grover's operators, and $O(1)$ measurements.

**Proof:**

It is indicated from **Algorithm 4-3** that Step (1) is the only loop and includes $n$ iterations to execute those steps embedded in the only loop. Hence, on the first execution of Step (1a), it calls for **Algorithm 4-2**. It is pointed out from Step (1) of **Algorithm 4-2** that $(n + 1)$ **Hadamard** gates are completed. Next, it is indicated from Step (2) of **Algorithm 4-2** that $(2 \times m)$ **CCNOT** gates are performed. It is then indicated from **Algorithm 4-2** that Step (4a) is embedded in the first nested loop and results in $(\dfrac{n \times (n+1) \times (n+2)}{3})$ **NOT** gates and $(\dfrac{n \times (n+1) \times (n+14)}{6})$ **CCNOT** gates. Next, it is pointed out from Step (5) of **Algorithm 4-2** that one **CNOT** gate is finished. It is then indicated from **Algorithm 4-2** that Step (6) of **Algorithm 4-2** is applied to restore the auxiliary quantum bits back to their original status. Thus, it is pointed out from **Algorithm 4-2** that Step (6) results in $(\dfrac{n \times (n+1) \times (n+2)}{3})$ **NOT** gates and $((2 \times m) + (\dfrac{n \times (n+1) \times (n+14)}{6}))$ **CCNOT** gates. This is to say that Steps (2) through (6) are employed to complete the oracle work. It is indicated from Step (7) of **Algorithm 4-2** that one Grover's operator in **Grover's algorithm** is performed. From Step (8) of **Algorithm 4-2** and **Grover's algorithm**, $(\sqrt{2^n})$ oracle works and $(\sqrt{2^n})$ **Grover's** operators are obtained. In Step (9) of **Algorithm 4-2**, a measurement is completed. Therefore, after the first call of **Algorithm 4-2** is performed, it is derived that O($n + 1$) **Hadamard** gates, O($2^{\frac{n}{2}} \times (\frac{2 \times n^3 + 6 \times n^2 + 4 \times n}{3})$) **NOT** gates, O($2^{\frac{n}{2}}$) **CNOT** gates, O($2^{\frac{n}{2}} \times (4 \times m + (\frac{n^3 + 15 \times n^2 + 14 \times n}{3}))$) **CCNOT** gates, O($2^{\frac{n}{2}}$) Grover's operators, and O(1) measurements are obtained.

Next, after the first call of **Algorithm 4-2** is completed, in the first execution of Step (1b) in **Algorithm 4-3**, and if the answer is obtained in the first execution of Step (1a) in **Algorithm 4-3**, then **Algorithm 4-3** is terminated in the first execution of Step (1c) in **Algorithm 4-3**. Therefore, the best case for time complexity of solving an instance of the clique problem to any graph $G = (V, E)$ with $n$ vertices and $\theta$ edges and its *complementary* graph $\overline{G} = (V, \overline{E})$ with $n$ vertices and $m = \frac{n \times (n-1)}{2} - \theta$ edges is O($n + 1$) **Hadamard** gates, O($2^{\frac{n}{2}} \times (\frac{2 \times n^3 + 6 \times n^2 + 4 \times n}{3})$) **NOT** gates, O($2^{\frac{n}{2}}$) **CNOT** gates, O($2^{\frac{n}{2}} \times (4 \times m + (\frac{n^3 + 15 \times n^2 + 14 \times n}{3}))$) **CCNOT** gates, O($2^{\frac{n}{2}}$) Grover's operators, and O(1) measurements. ∎

**Lemma 5-2**: The worst case for time complexity of solving an instance of the clique problem for any graph $G = (V, E)$ with $n$ vertices and $\theta$ edges and its *complementary* graph $\overline{G} = (V, \overline{E})$ with $n$ vertices and $m = \frac{n \times (n-1)}{2} - \theta$ edges is O($n \times (n + 1)$) **Hadamard** gates, O($n \times (2^{\frac{n}{2}} \times (\frac{2 \times n^3 + 6 \times n^2 + 4 \times n}{3}))$) **NOT** gates, O($n \times 2^{\frac{n}{2}}$) **CNOT** gates, O($n \times (2^{\frac{n}{2}} \times (4 \times m + (\frac{n^3 + 15 \times n^2 + 14 \times n}{3})))$) **CCNOT** gates, O($n \times (\sqrt{2^n})$) Grover's operators, and O($n \times 1$) measurements.

**Proof:**

It is indicated from **Algorithm 4-3** that the worst case of solving an instance of the clique problem for any graph $G = (V, E)$ with $n$ vertices and $\theta$ edges and its *complementary* graph $\overline{G} = (V, \overline{E})$ with $n$ vertices and $m = \frac{n \times (n-1)}{2} - \theta$ edges is finding the answer after a measurement on the result generated from the $n$th execution of Step (1a) in **Algorithm 4-3** is completed. This is to say that each step in **Algorithm 4-3** is executed $n$ times. Therefore, the worst case of time complexity is O($n \times (n + 1)$) **Hadamard** gates, O($n \times (2^{\frac{n}{2}} \times (\frac{2 \times n^3 + 6 \times n^2 + 4 \times n}{3}))$) **NOT** gates, O($n \times 2^{\frac{n}{2}}$) **CNOT** gates, O($n \times (2^{\frac{n}{2}} \times (4 \times m + (\frac{n^3 + 15 \times n^2 + 14 \times n}{3})))$) **CCNOT** gates, O($n \times (\sqrt{2^n})$) Grover's operators, and O($n \times 1$) measurements. ∎

**Lemma 5-3**: The average case for time complexity of solving an instance of the clique problem for any graph $G = (V, E)$ with $n$ vertices and $\theta$ edges and its *complementary* graph $\overline{G} = (V, \overline{E})$ with $n$ vertices and $m = \frac{n \times (n-1)}{2} - \theta$ edges is O($(\frac{n+1}{2}) \times (n+1)$) **Hadamard** gates, O($(\frac{n+1}{2}) \times (2^{\frac{n}{2}} \times (\frac{2 \times n^3 + 6 \times n^2 + 4 \times n}{3}))$)

NOT gates, $O((\frac{n+1}{2}) \times 2^{\frac{n}{2}})$ CNOT gates, $O((\frac{n+1}{2}) \times (2^{\frac{n}{2}} \times (4 \times m + (\frac{n^3 + 15 \times n^2 + 14 \times n}{3}))))$ CCNOT gates, $O((\frac{n+1}{2}) \times (\sqrt{2^n}))$ Grover's operators, and $O(\frac{n+1}{2})$ measurements.

**Proof:**

It is supposed that the time complexity of one time for executing each step in **Algorithm 4-3** is $\Omega$. Hence, the average case of time complexity is $((1 \times \Omega) + (2 \times \Omega) + \ldots + (n \times \Omega)) \div (n) = (\frac{n+1}{2}) \times \Omega$ because $\Omega$ is equal to the best case of time complexity. Therefore, it is derived that the average case of time complexity is $O((\frac{n+1}{2}) \times (n+1))$ **Hadamard** gates, $O((\frac{n+1}{2}) \times (2^{\frac{n}{2}} \times (\frac{2 \times n^3 + 6 \times n^2 + 4 \times n}{3})))$ **NOT** gates, $O((\frac{n+1}{2}) \times 2^{\frac{n}{2}})$ **CNOT** gates, $O((\frac{n+1}{2}) \times (2^{\frac{n}{2}} \times (4 \times m + (\frac{n^3 + 15 \times n^2 + 14 \times n}{3}))))$ **CCNOT** gates, $O((\frac{n+1}{2}) \times (\sqrt{2^n}))$ Grover's operators, and $O(\frac{n+1}{2})$ measurements. ∎

**Lemma 5-4**: The best case for space complexity of solving an instance of the clique problem for any graph $G = (V, E)$ with $n$ vertices and $\theta$ edges and its *complementary* graph $\overline{G} = (V, \overline{E})$ with $n$ vertices and $m = \frac{n \times (n-1)}{2} - \theta$ edges is $O((2 \times m + 3) + (\frac{n^3 + 15 \times n^2 + 26 \times n}{6}))$ quantum bits.

**Proof:**

Since there are $2^n$ possible choices (including all possible cliques), a quantum register with $n$ bits $(\otimes_{b=n}^{1} |x_b\rangle)$ is used to encode $2^n$ choices with initial states $(\otimes_{b=n}^{1} |x_b^0\rangle)$. The clique problem of any graph $G = (V, E)$ with $n$ vertices and $\theta$ edges and its *complementary* graph $\overline{G} = (V, \overline{E})$ with $n$ vertices and $m = \frac{n \times (n-1)}{2} - \theta$ edges requires finding a maximum-sized clique from $G$, so those auxiliary quantum registers are needed. The initial states of those auxiliary quantum registers are $(|1\rangle) \otimes (\otimes_{i=n}^{1} \otimes_{j=i}^{0} |z_{i,j}^0\rangle) \otimes (|z_{0,0}^1\rangle) \otimes (\otimes_{i=n-1}^{0} \otimes_{j=i}^{0} |g_{i,j}^0\rangle) \otimes (\otimes_{i=n-1}^{0} \otimes_{j=i}^{0} |f_{i,j}^0\rangle) \otimes (\otimes_{i=n-1}^{0} \otimes_{j=i}^{0} ((\otimes_{a=i-j+1}^{1} |h_{i,j,a}^0\rangle) \otimes (|h_{i,j,0}^1\rangle))) \otimes (\otimes_{k=m}^{1} |c_k^0\rangle) \otimes (|c_0^1\rangle) \otimes (\otimes_{k=m}^{1} |r_k^1\rangle) \otimes (\otimes_{b=n}^{1} |x_b^0\rangle)$. Therefore, it is pointed out from **Algorithm 4-3** that the best case of space complexity is to find the answer after **Algorithm 4-2** is only invoked once. Hence, the best case of space complexity is $O((2 \times m + 3) + (\frac{n^3 + 15 \times n^2 + 26 \times n}{6}))$ quantum bits. ∎

**Lemma 5-5**: The worst case for space complexity of solving an instance of the clique problem for any graph $G = (V, E)$ with $n$ vertices and $\theta$ edges and its *complementary* graph $\overline{G} = (V, \overline{E})$ with $n$ vertices and $m =$ $\frac{n \times (n-1)}{2} - \theta$ edges is $O(n \times ((2 \times m + 3) + (\frac{n^3 + 15 \times n^2 + 26 \times n}{6})))$ quantum bits.

**Proof:**

From **Lemma 5-2** and **Algorithm 4-3**, it is inferred that the worst case of space complexity is $O(n \times ((2 \times m + 3) + (\frac{n^3 + 15 \times n^2 + 26 \times n}{6})))$ quantum bits. ∎

**Lemma 5-6**: The average case for space complexity of solving an instance of the clique problem for any graph $G = (V, E)$ with $n$ vertices and $\theta$ edges and its *complementary* graph $\overline{G} = (V, \overline{E})$ with $n$ vertices and $m = \frac{n \times (n-1)}{2} - \theta$ edges is $O((\frac{n+1}{2}) \times ((2 \times m + 3) + (\frac{n^3 + 15 \times n^2 + 26 \times n}{6})))$ quantum bits.

**Proof:**

From **Lemma 5-3** and **Algorithm 4-3**, it is concluded that the average case for space complexity is $O((\frac{n+1}{2}) \times ((2 \times m + 3) + (\frac{n^3 + 15 \times n^2 + 26 \times n}{6})))$ quantum bits. ∎

## 6. CONCLUSIONS

From [Ho et al. 2005], it is indicated that the DNA-based algorithm for solving an instance of the clique problem for any graph $G = (V, E)$ with $n$ vertices and $\theta$ edges and its *complementary* graph $\overline{G} = (V, \overline{E})$ with $n$ vertices and $m = \frac{n \times (n-1)}{2} - \theta$ edges includes four main steps. From the first phase, $2^n$ DNA strands is employed to represent $2^n$ combinations states that can be implemented by means of arbitrary superposition of $n$ quantum bits ($n$ **Hadamard** gates operating on $n$ quantum bits). From the second phase, computing legal cliques and removing illegal cliques can be implemented by **CCNOT** gates (the quantum circuit in Figure 4-7). From the third phase, determining the number of vertices for each legal clique can be also implemented by **NOT** gates and **CCNOT** gates (the quantum circuit in Figure 4-8). From the fourth phase, reading a maximum-sized clique that contains the maximum vertices can be implemented by **Grover's algorithm**. This is to say from **Lemma 4-7** that if **Grover's algorithm** is used to complete the readout step in the DNA-based algorithm, the quantum implementation of the DNA-based algorithm is equivalent to the oracle work (in the language of **Grover's algorithm**), that is, the target state labeling preceding **Grover's** searching steps. Quantum implementation of the DNA-based algorithm assures having a faster labeling of the target state, which also implies a speedy solution for an instance of the clique problem of any graph $G = (V, E)$ with $n$ vertices and $\theta$ edges and its *complementary* graph $\overline{G} = (V, \overline{E})$ with $n$ vertices and $m = \frac{n \times (n-1)}{2} - \theta$ edges. The experimental results are in well agreement with the theoretical prediction. Although the current technology is only available for the simplest case of the clique problem, we believe, with more advanced NMR technique and more experimental efforts, more complicated clique solutions could be demonstrated by NMR experiments. We also expect that our algorithm could be carried out by a real quantum machine in the future.

## ACKNOWLEDGEMENTS

This work is partly supported by the National Natural Science Foundation of China under Grants No. 10774163, partly by the National Fundamental Research Program of China under Grants No. 2006CB921203, and also partly supported by the National Science Foundation of Republic of China under Grants No. 96-2221-E-151-008-,

96-2218-E-151-004- and 97-2221-E-151-035-.REFERENCE

Adleman, L. 1994. Molecular computation of solutions to combinatorial problems. *Science*, 266: No. 11, 1021-1024.

Adleman, L., Rothemund, Paul W. K., Roweis, S., and Winfree, E.. 1999. On applying molecular computation to the data encryption standard. *The 2nd annual workshop on DNA Computing*, Princeton University, DIMACS: series in Discrete Mathematics and Theoretical Computer Science, American Mathematical Society, 31-44.

Ambainis, A. and Spalek, R. 2006. Quantum algorithms for matching and network flows. *Proceedings of STACS'06, Lecture Notes in Computer Science*, Volume 3884/2006, 172-183.

Bernstein, E. and Vazirani, U. 1993. Quantum complexity theory. *Proceedings of the twenty-fifth annual ACM symposium on Theory of computing*, 11-20.

Braich, R. S., Johnson, C., Rothemund, P.W. K., Hwang, D., Chelyapov, N. and Adleman, L. M. 2002. Solution of a 20-variable 3-SAT problem on a DNA computer. *Science*, vol. 296, No. 5567, 499–502.

Boneh, D., Dunworth, C and Lipton, R. J. 1996. Breaking DES using a molecular computer. In *Proceedings of the 1st DIMACS Workshop on DNA Based Computers*, 1995, American Mathematical Society. In *DIMACS Series in Discrete Mathematics and Theoretical Computer Science,* Volume 27, 37-66.

Benioff, P. 1982. Quantum mechanical models of turing machines. .*Physical Review Letter*, 48, 1581-1585.

Bennett, C. H., Bernstein, E., Brassard, G. and Vazirani, U. V. 1997. Strengths and weakness of quantum computing. *SIAM Journal on Computing*, Volume 26, No. 5, 1510-1523.

Berzina, A., Dubrovsky, A., Freivalds, R., Lace, L. and Scegulnaja1,O. 2004. Quantum query complexity for some graph problems. *Lecture Notes in Computer Science*, Volume 2932/2003, 140-150.

Cook, S. 1971. The complexity of theorem proving procedures. *Proceedings of Third Annual ACM Symposium on Theory of Computing*, 151-158.

Chang, W.-L., Ho, M., and Guo, M. 2004. Molecular solutions for the subset-sum problem on DNA-based supercomputing. *BioSystems*, Volume 73, No. 2, 117-130.

Chang, W.-L., Ho, M. and Guo, M. 2005. Fast parallel molecular algorithms for DNA-based computation: factoring integers. *IEEE Transactions on Nanobioscience*, Vol. 4, No. 2, 149-163.

Coppersmith, D. 1994. An approximate Fourier transform useful in quantum factoring. *IBM research report, Technical Report RC 19642,* IBM Research Division T.J. Watson Research Center, e-print quantph/0201067.

Cory, D. G., Price, M. D., Havel, T. F., 1998. Nuclear magnetic resonance pectroscopy: an experimentally accessible paradigm for quantum computing. *Phyisca D*, 120 (1-2): 82-101.

Chuang, I. L., Gershenfeld, N. M., Kubinec, G. and Leung, D. W. 1998. Bulk quantum computation with nuclear magnetic resonance: theory and experiment, *Proc. R. Soc. Lond*. 454: 447-455.

Debnath, L. and Mikusinski, P. 1990. *Introduction to Hilbert Spaces with Applications*. Academic Press, INC, **ISBN** 0-12-208435-7.

Deutsch, D. 1985. Quantum theory, the Church-Turing principle and the universal quantum computer. *Proceeding of Royal Society of London series A*, 400, 97-117.

Deutsch, D. 1989. Quantum computational networks. *Proceedings of the Royal Society of London, Series A, Mathematical and Physical Sciences*, Volume 425, No. 1868, 73-90.

Deutsch, D. and Jozsa, R. 1992. Rapid solutions of problems by quantum computation. *Proceedings of the Royal Society of London, Series A*, Vol. 439, 553-558.

Doern, S. 2007. Quantum algorithms and lower bounds for independent set and subgraph isomorphism problem. *Proceedings of SOFSEM'07*, quant-ph/0510084

Durr, C., Heiligman, M., Hoyer, P. and Mhalla, M. 2004. Quantum query complexity of some graph problems. *Proceedings of ICALP'04*: 481-493.

Ernst, R. R., Bodenhausen, G., Wokaun, A. 1987.*Principles of Nuclear Magnetic Resonance in One and Two Dimensions,* Oxford Press.

Feynman, R. P. 1961. In minaturization. D.H. Gilbert, Ed., Reinhold Publishing Corporation, New York, 282-296.

Feynman, R. P. 1982. Simulating physics with computers. *Internation Journal of Theoretical Physics*, Volume 21, No. 6/7, 467-488.

Garey, M. R., and Johnson, D. S. 1979. *Computer and Intractability*: *A Guide to the Theory of NP-Completeness*. W. H. Freeman Company, New York, **ISBN** 0716710447.

Guo, M., Chang, W.-L., Ho, M., Lu, J. and Cao, J. 2005. Is optimal solution of every NP-Complete or NP-Hard problem determined from its characteristic for DNA-based computing. *BioSystems*, Vol. 80, No. 1, 71-82.

Grover, L. K. 1996. A fast quantum mechanical algorithm for database search. *Proceedings of the twenty-eighth annual ACM symposium on Theory of computing*, 212-219.

Ho, M., Chang, W.-L., Guo, M., and Yang, T. L. 2004. Fast parallel solution for set-packing and clique problems by DNA-based Computing. *IEICE Transactions on Information and Systems*, Volume E-87D, No. 7, 1782-1788.

Hogg, T. 1998 .Highly structured searches with quantum computers. *Phys. Rev. Lett*. 80, 2473 – 2476.

Hopcroft, J. E., Motwani, R. and Ullman, J. D. 2001. *Introduction to Automata Theory, Languages, and Computation*, Addison Wesley, **ISBN** 0-201-44124-1.27